%% file: main.tex
\documentclass[
 reprint,
superscriptaddress,
 amsmath,
 amssymb,
 aps,
 pra,
showkeys
]{revtex4-2}

\usepackage{graphicx}
\usepackage{dcolumn}
\usepackage{bm}
\usepackage{booktabs}
\usepackage{tabularx}
\usepackage[separate-uncertainty=true]{siunitx}
\DeclareSIUnit\solarmass{M\ensuremath{_\odot}}
\DeclareSIUnit\year{yr}
\DeclareSIUnit\parsec{pc}
\usepackage{comment}
\usepackage{aas_macros}

\usepackage{xcolor}

\usepackage[normalem]{ulem}

\usepackage{xcolor}
\definecolor{darkblue}{rgb}{0.0, 0.0, 0.55}
\usepackage[colorlinks=true, allcolors=darkblue]{hyperref}
\usepackage{orcidlink}

\begin{document}

\preprint{APS/123-QED}

\title{Assessing the performance of future space-based detectors:\\Astrophysical foregrounds and individual sources}

\author{Alice Perego \orcidlink{0009-0001-0670-2738}} \email[Contact author: ]{alice.perego@oca.eu}
\affiliation{Université Côte d’Azur, Observatoire de la Côte d’Azur, Laboratoire Artemis, CNRS, Bd de l’Observatoire, 06300 Nice, France}
\affiliation{Université Côte d’Azur, Observatoire de la Côte d’Azur, Laboratoire Lagrange, CNRS, Bd de l’Observatoire, 06300 Nice, France}

\author{Matteo Bonetti \orcidlink{0000-0001-7889-6810}}
\affiliation{
Dipartimento di Fisica “G. Occhialini”, Università degli Studi di Milano-Bicocca, Piazza della Scienza 3, 20126 Milano, Italy}
\affiliation{
INFN - Sezione di Milano-Bicocca, Piazza della Scienza 3, 20126 Milano, Italy
}

\author{Alberto Sesana \orcidlink{0000-0003-4961-1606}}
\affiliation{
Dipartimento di Fisica “G. Occhialini”, Università degli Studi di Milano-Bicocca, Piazza della Scienza 3, 20126 Milano, Italy}
 \affiliation{
 INFN - Sezione di Milano-Bicocca, Piazza della Scienza 3, 20126 Milano, Italy
 }
 
\author{Silvia Toonen \orcidlink{0000-0002-2998-7940}}
\affiliation{
 Anton Pannekoek Institute for Astronomy, University of Amsterdam, 1090 GE, Amsterdam, the Netherlands
 }

\author{Valeriya Korol \orcidlink{0000-0002-6725-5935} }
\affiliation{SRON Space Research Organisation Netherlands, Niels Bohrweg 4, 2333 CA Leiden, the Netherlands}
\affiliation{Max Planck Institute for Astrophysics, Karl-Schwarzschild-Straße 1, 85748 Garching, Germany}

\date{\today}

\begin{abstract}
The space mission LISA, scheduled for launch in 2035, aims to detect gravitational wave (GW) signals in the milli-Hz band.
In the context of the ESA \textit{Voyage 2050} Call for new mission concepts, other frequency ranges are explored by the Gravitational-Wave Space 2050 Working Group to conceive new proposals for a post-LISA space-based detector. In this work, we give a preliminary estimate of the observational potential of three mission designs proposed in the literature, namely \textit{$\mu$Ares}, \textit{AMIGO} and the \textit{Decihertz Observatory}. The analysis framework includes astrophysical GW sources, such as massive black hole binaries and extreme mass-ratio inspirals, and compact binaries, such as stellar black holes and white dwarfs. For each detector, we first present a consistent computation of the unresolved gravitational wave background (GWB) produced by the sum of all anticipated astrophysical populations using an iterative subtraction algorithm. We then investigate which types of systems are the most appealing by measuring the number of GW signals detected and exploring the source properties.
\end{abstract}

\maketitle

\input{introduction}

\input{detectors}

\input{sources}
\input{methods}
\input{results}
\input{discussion_conclusions}

\bibliography{bibliography}

\end{document}

%% file: introduction.tex
\section{Introduction}

Until the last decade, the majority of astrophysical observations were made by telescopes in the electromagnetic (EM) spectrum, from the radio to the gamma-ray band. Cosmic rays and neutrinos added to the mix, offering a window on the physics of energetic processes occurring in the Universe. On September 2015, the gravitational wave (GW) signal emitted by a stellar-origin binary black hole (SOBBH) merger \cite{GW150914} was detected for the first time: it marked the beginning of gravitational-wave astronomy, opening a completely new window on the Universe. As for EM waves, GWs also span a broad array of frequencies and are generated by an equally wide variety of sources.

The first observatory to enter in operation was the Laser Interferometer Gravitational-Wave Observatory (LIGO) \cite{Advanced_LIGO}, whose advanced version started observing in 2015. It was later joined in 2017 by the Advanced Virgo interferometer \cite{Virgo}, located in Cascina, Italy, and in 2020 by the Kamioka Gravitational Wave Detector (KAGRA) \cite{KAGRA}, located in Gifu Prefecture, Japan. They initiated a global network of GW observatories targeting GWs in the frequency range from $\SI{10}{\hertz}$ to a few $\SI{}{\kilo \hertz}$,populated by the signal of merging binaries formed of two closely orbiting compact objects, like stellar-mass black holes (BHs) or neutron stars (NSs).
Given the success of these observations, a third generation of ground-based detectors are currently under study. The European Einstein Telescope (ET) \cite{Punturo_2010_ET} and the US-based Cosmic Explorer \cite{Reitze_2019_CE} are experiments aiming at building interferometers with greater arm length and improved technology, which will increase the instrumental sensitivity and extend the observable range of GW frequencies below $\SI{10}{\hertz}$.

On the opposite side of the GW frequency spectrum, a different technique is employed to probe the nano-Hz band. The detection of GWs in this window is the primary goal of the Pulsar Timing Array (PTA) community, which exploits the largest radio telescopes to monitor an array of tens of millisecond pulsars. This effectively creates a galactic size detector that aims at detecting tiny correlations patterns in the time of arrival of the radio pulses \cite{1978SvA....22...36S}. In June 2023, several PTA collaborations independently announced that they found evidence for a stochastic gravitational wave background (GWB) in the frequency range \SI{10^{-9}-10^{-6}}{\hertz} \cite{EPTA,NANOGrav,PPTA,CPTA}, with 2-to-4$\sigma$ confidence depending on the dataset. The most likely candidate for this signal is the cosmic population of orbiting pairs of supermassive black holes (SMBHs) \cite{PTA_Astro, EPTA_2, 2023ApJ...952L..37A}, which collectively produce a superposition of GW signals in the nano-Hz band.

Between the nano-Hz frequency band targeted by PTA and the $\SI{10-10^3}{\hertz}$ range where LVK operates lies the unexplored milli-Hz window. The planned Laser Interferometer Space Antenna (LISA) will cover this gap, revealing GW signals between $\SI{0.1~-~100}{\milli \hertz}$ \cite{LISA_red_book}. This space interferometer was selected as one of the three large-class missions of the campaign \textit{Cosmic Vision 2015-2025} and the project was formally adopted in 2024 \cite{LISA_adoption}. The implementation phase has begun, with the goal of launching the mission in 2035. The milli-Hz frequency band is anticipated to be the richest in terms of variety of astrophysical sources and LISA will detect massive black hole binaries (MBHBs), extreme mass-ratio inspirals (EMRIs) and compact binaries, primarily composed of double white dwarfs (DWDs). It will provide answers to numerous intriguing scientific questions, from astrophysics to cosmology and fundamental physics, comprehensively detailed in \citet{LISA_astro}. In addition to LISA, the milli-Hz range will also be explored by two other space-based detectors, the Chinese projects Taiji \cite{Taiji_2017} and TianQin \cite{TianQin_2016}, both expected to operate in the 2030s.

The upcoming European Space Agency (ESA) science program campaign \textit{Voyage 2050} has already been defined, and it will outline the space science missions for the period 2035-2050 \cite{Favata2021}. One of the three scientific areas selected focuses on \textit{New Physical Probes of the Early Universe}, through the cosmic microwave background (CMB) or GW signals. Within the Gravitational-Wave Space 2050 Working Group, operating in line with the perspective offered by Voyage 2050, new concepts for a post-LISA space-based detector have been proposed, exploring frequency regions beyond the milli-Hz.

Previous work have already identified the scientific outcome of the milli-Hz window \citep{Sesana2021muAres, Martens2023lisamax} and the deci-Hz band \citep{Sedda2020decihertz}. Despite being comprehensive in analyzing the number of detection per kind of sources, a crucial missing piece in those studies was a systematic investigation of astrophysical foreground signals generated from the GW sources and possibly affecting the actual detection of single events. This issue is particularly relevant for the frequency range from $\SI{10^{-6}}{\hertz}$ to $\SI{1}{\hertz}$, which is expected to be signal-dominated, due to the high number of astrophysical sources emitting simultaneously in this window. The largest fraction of these signals will be below the detection threshold of space interferometers, thus producing a cumulative stochastic signal that can limit the detector sensitivity, if exceeding the instrumental noise \cite{Regimbau_GWB}. For these reasons, estimating and modeling astrophysical gravitational-wave backgrounds (GWBs) have been the focus of research in multiple studies, especially in the context of LISA and PTA. Their detection and reconstruction is also a challenging task, not only because they have to be distinguished from the detector noise, but also from stochastic GWBs of cosmological origin \cite{Romano_GWB,Caprini_GWB, Renzini}.

Especially relevant for the LISA mission is the population of DWDs within our galaxy. The majority of them will produce an anisotropic overall signal, known as \textit{Galactic foreground}, that will affect LISA's observations up to a few $\SI{}{\milli \hertz}$. Its characterization is a key factor in LISA data analysis; therefore, various approaches have been explored. A Bayesian strategy, including Markov Chain Monte Carlo algorithms and model selection methods, can be a promising option \cite{Littenberg,Boileau_WD_GWB}, despite its computational expense. An alternative solution is the computation of the global unresolved GWB using an iterative procedure for the subtraction of the detectable sources, as described in \citet{Karnesis_code} (for related works, see also \cite{2006PhRvD..73l2001T, 2007CQGra..24S.575C, 2011PhRvD..84f3009L, 2012ApJ...758..131N}). This method, used in \citet{Korol_WD} to model the Galactic foreground and in \citet{Torrado_SOBBH} for the GWB from SOBBHs, is more straightforward as it assumes idealized detector noise and perfect source subtraction, while still providing a valuable estimate of the stochastic signal. Other astrophysical populations expected to produce a GWB relevant for space-based interferometers are EMRIs \cite{Pozzoli_EMRI}, MBHBs \cite{Sesana_Vecchio_Colacino} and extragalactic compact binaries such as DWDs \cite{Phinney_extragal, Staelens_Nelemans} and SOBBHs \cite{Torrado_SOBBH}.

In this paper, we present a self-consistent computation of astrophysical GWBs from the source populations listed above, carried out for the first time within a unified framework that relies on the source-subtraction algorithm of \citet{Karnesis_code}. We demonstrate the versatility of the calculation by applying it on a number of different detectors between the one proposed within Voyage2050. For each of them, we compute the cumulative unresolved GWB, incorporate it into the instrumental noise, and assess its impact on the detection of individual sources, ultimately evaluating the overall performance of the mission in terms of astrophysical science return.

The paper is organized as follows. Section~\ref{sec:missions} introduces the different mission designs, while Section~\ref{sec:sources} describes the astrophysical populations considered. In Section~\ref{sec:methods} we detail the calculation for the signal-to-noise ratio (SNR) and the GWB, along with the structure of the iterative code we employed. Section~\ref{sec:results} presents the results for both the unresolved GWB and the resolved sources. Finally, in Section~\ref{sec:discussion_conclusions} we discuss the main results and draw our conclusions.

%% file: detectors.tex
\section{Mission concepts}
\label{sec:missions}

In this study, we evaluate the performance of three proposed detector configurations, \textit{$\mu$Ares} from \citet{Sesana2021muAres}, the \textit{Decihertz Observatory (DO)} from \citet{Sedda2020decihertz}, and the \textit{Advanced Millihertz Gravitational-wave Observatory (AMIGO)} from \citet{Baibhav2021Amigo}, comparing the results with LISA capability. All of them are space-based interferometers and the detection principle is the same as the one employed by LISA: 
three identical satellites exchange laser beams with each other in order to monitor the relative separation of the free fall test masses contained in each of them. The main difference lies in the orbits of the three spacecrafts, which determine the size and the motion of the triangular constellation.

We describe the performance of each detector via its sky-averaged noise power spectral density (PSD), $S_n(f)$ (where $f$ is the observed frequency), which is the PSD of the noise, multiplied by a numerical factor that takes into account the response of the detector averaged over the incoming direction of the source. Regardless of the specific orbit, for an equilateral triangle and two independent channels, this factor is 10/3, and the averaged PSD takes the form:

\begin{equation}
    S_n(f) = \frac{10}{3} \left [ A\,\frac{1 + \left ( \frac{f_1}{f}\right)^2}{(2\pi f)^4} + B \right] \times \left[1 + \left( \frac{f}{f_2}\right)^2 \right]
\label{eq:PSD}
\end{equation}

Equation~\eqref{eq:PSD} includes a low frequency test mass acceleration noise component described by the first term in the first square bracket, and a high frequency noise component due to the optical metrology system, encapsulated in the second square bracket. The transition between the different noise regimes is set by the critical frequencies $f_1$ and $f_2$, and the noise level is defined by the dimensional constants $A$ [s$^{-4}$Hz$^{-1}$] and $B$ [Hz$^{-1}$]. The values of these quantities for all the detectors considered is given in Tab. \ref{tab:noise}.

\begin{table}[t]
\begin{ruledtabular}
\begin{tabular}{c|c|c|c|c}
\textrm{ }&
A [s$^{-4}$Hz$^{-1}$]&
B [Hz$^{-1}$]&
$f_1$ Hz&
$f_2$ Hz\\
\colrule
LISA & $5.76\times10^{-48}$ & $3.6\times10^{-41}$ & $4\times10^{-4}$ & $2.5\times10^{-2}$ \\
AMIGO & $5.76\times10^{-50}$ & $3.6\times10^{-43}$ & $4\times10^{-4}$ & $2.5\times10^{-2}$ \\
$\mu$Ares & $5.13\times10^{-53}$ & $1.7\times10^{-44}$ & $2\times10^{-6}$ & $1.5\times10^{-4}$ \\
DO & $3.60\times10^{-47}$ & $3.6\times10^{-45}$ & $4\times10^{-4}$ & $6.1\times10^{-2}$ \\
\end{tabular}
\end{ruledtabular}
\caption{\label{tab:noise}
Parameters entering Eq.~\eqref{eq:PSD} for each of the four detectors investigated in this study. The resulting sensitivity curves are shown in Fig.~\ref{fig:ASD_inst}. }

\end{table}

LISA has an arm length of $\SI{2.5 \times 10^6}{\kilo \meter}$, and the parameters that produce its PSD are provided  by the Science Requirement Document \cite{SciRD}.

AMIGO \cite{Baibhav2021Amigo} retains the same spatial configuration and placement as LISA, employing interferometer arms of $\SI{2.5 \times 10^6}{\km}$, while introducing a tenfold enhancement in strain sensitivity across all frequencies. Therefore, it retains the same $f_1$ and $f_2$, featuring a $\times100$ improvement in $A$ and $B$ (cf. Tab. \ref{tab:noise}, strain sensitivity is related to the square root of the PSD).

The design of the $\mu$Ares mission \cite{Sesana2021muAres} is rather different from the LISA one. It consists of three spacecrafts with an orbit radius of 1.5 AU, resulting in a triangular constellation centered in the Sun with an average arm length of $\SI{395 \times 10^6}{\km}$, more than two orders of magnitude longer than LISA's arms. This affects the number of photons received at the mirrors, shifting $f_1$ and $f_2$ to much lower values and making the instrument sensitive down to $\mu$-Hz.

The design of DO \cite{Sedda2020decihertz} resembles that of LISA, featuring a triangular configuration in an Earth-trailing orbit, but with an arm length 25 times shorter ($\SI{10^5}{\km}$) which enhances the sensitivity at higher frequencies. We will employ the less ambitious version of the various deci-Hz mission concepts, the DO-Conservative.

\begin{figure}[b]
    \includegraphics[width=\hsize]{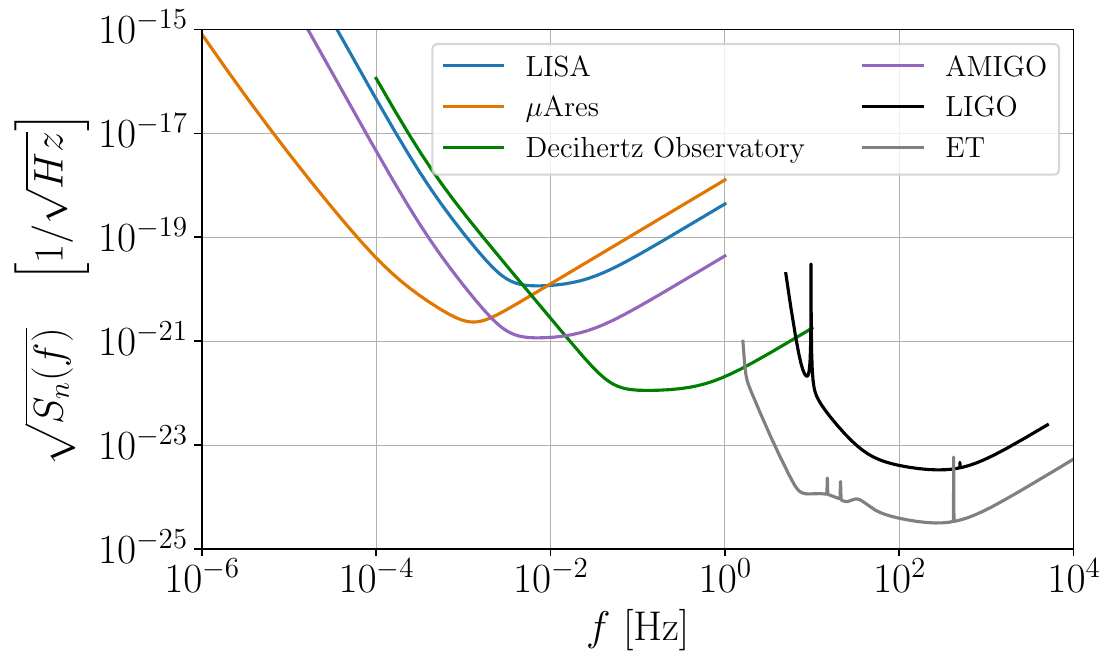}
    \caption{ Amplitude spectral density of the sensitivity curves for the four detectors.}
    \label{fig:ASD_inst}
\end{figure}

The square root of the sky averaged PSD (also referred to as Amplitude Spectral Density -- ASD) for each of the four space-based detectors is illustrated in Fig. \ref{fig:ASD_inst}, compared to current and next-generation terrestrial interferometers LIGO and ET.

Each detector is also characterized by its mission duration: we consider a science phase of $T_{\mathrm{obs}}$ = 10 yr for all the missions. During this time no time-variations of the PSD due to glitches \cite{Baghi_glitches}, gaps \cite{Baghi_gaps} or the detector motion \cite{Digman_2022} are included in the analysis. We will comment on the impact of these assumptions in Sec.~\ref{sec:caveats}.

%% file: sources.tex
\section{Source catalogues}
\label{sec:sources}
In this study we will consider these five classes of GW sources: MBHBs, EMRIs, SOBBHs, Galactic binaries (GBs) and extragalactic DWDs. We neglect the possible contribution to the GW source budget brought by intermediate mass black holes (IMBHs) possibly formed in globular or star clusters \cite{2002MNRAS.330..232C,2004Natur.428..724P}. While merger of two IMBHs following cluster mergers might be rare, IMBHs can capture stellar compact objects from their parent cluster \citep[e.g.][]{2010ApJ...719..987M} or fall into galactic centers pairing with the central MBH \citep[e.g.][]{2006ApJ...641..319P}. They can therefore be either the primary or the secondary component of intemediate mass ratio inspirals (IMRIs). Since the evidence of the existence of IMBHs is sparse \cite{2020ARA&A..58..257G}, we do not consider these IMRI populations here. We stress, however, that our framework is modular, and any astrophysical binary population can be trivially added when available.

\subsection{Massive black hole binaries}

In this work, we make use of the population catalogs employed in \citet{Bonetti_MBHB} in the context of the LISA mission. The coevolution of massive black holes (MBHs) and their host galaxies is simulated using the semi-analytical model of \citet{Barausse_SAM}, with improvements from subsequent studies. For the birth mechanism of the MBHs, two different prescriptions are adopted \cite{Klein_MBHB}. In the light-seed (LS) scenario, the MBH seeds are remnants of population-III stars at high redshift ($z \sim 15-20$), and they populate the center of the most massive halos only. In contrast, the heavy-seed (HS) model generates MBH seeds through the collapse of proto-galactic discs caused by bar instabilities, resulting in seeds that are already massive at high redshift. 

We employ two catalogs, corresponding to the HS and LS models, which are characterized by different mass function. Basically, for the LS model the bulk of the population is concentrated at high redshift (up to $\sim$ 20) with the low mass end reaching $\sim \SI{10^{2}}{\solarmass}$. Given the very low mass, the rest-frame GW frequency can be as high as $\SI{10^{-2}}{\hertz}$. In the HS scenario, the mass function is more sharply peaked at values around $\SI{10^4-10^5}{\solarmass}$ at moderately high redshift ($z < 10$). In order to adequately cover the measurement frequency band of $\mu$Ares, the MBHB population in both catalogs extends down to $\sim \SI{10^{-6}}{\hertz}$ in orbital frequency. The eccentricity is set to zero and spins are assumed to be aligned.
We will comment more on these assumptions in Sec.~\ref{sec:caveats}.

\subsection{Extreme mass-ratio inspirals}

We employ a catalog of EMRI events constructed from the populations of \citet{Babak_EMRI}, later exploited also in \citet{Bonetti_EMRI}. The features of the populations are strongly related to the astrophysical model assumed, which in turn relies on a great number of poorly known parameters as the mass function of the MBH, the properties of the inspiraling compact object and the ratio of successful EMRIs versus direct plunges. We consider model M1 of \citet{Babak_EMRI}, accepted as a fiducial model in between the most optimistic and pessimistic scenarios. It assumes a MBH population based on the LS prescription, with nearly maximal spins. The stellar cusp adheres to the $M-\sigma$ relation of \citet{Gultekin_Msigma}, which defines also the regrowth time of the cusp after its disruption due to MBHB mergers. The compact objects, whose mass is fixed to $\SI{10}{\solarmass}$, could also be pulled by the MBH into almost radial orbits, resulting in the loss of potential EMRI events. In this model the number of direct plunges per EMRI is set to 10. The maximum redshift set in the generation of the population is $z=4.5$. This was deemed sufficient by \citet{Babak_EMRI} to cover the LISA reach with ample margin, as we will see later, the DO reach can exceed this redshift. We will comment more on this aspect in Sec.~\ref{sec:resolved}.

\subsection{Galactic binaries}

Mock populations of GBs are predominantly developed through binary population synthesis methods, although observational-driven models are being explored too \cite{Korol_Toonen_WD}. In this work, we exploit the catalogs constructed from the population synthesis code \texttt{SeBa}, detailed in \citet{Toonen_SeBa}. The model adopts an evolutionary path that assumes the $\alpha$-formalism for the description of the common envelope phase \cite{Paczynski_1976, Webbink_1984}, based on the conservation of orbital energy. The catalogs are constructed assuming the Milky Way model of \citet{Toonen_Nelemans_2013} which provides, among other things, the Galactic positions of the binaries and the star formation rate. The models are therefore quantitatively equal to the $\alpha \alpha$ model in \citet{Korol_2017, Korol_2024}, with an extension to lower frequencies. Besides WD-WD binaries, the catalogs include also living stars as hydrogen-burning main sequence(MS) stars, forming MS-MS and WD-MS binaries. The orbit is assumed with zero eccentricity in the analysis. In order to reduce the computational cost, we undersample the binary population by a factor of 1:100 at GW frequencies below $\SI{0.5}{\milli \hertz}$. At higher frequencies, where the majority of resolvable sources are found, we retain the full catalog resolution to ensure more accurate calculations. We have tested that the undersampling does not affect neither the estimation of the unresolved GWB nor the recovery of individual sources.

\subsection{Stellar-origin binary black holes}

The merger rates and BHs properties inferred by the observation runs of LIGO and Virgo are frequently used to constrain the population parameters of compact binaries. In our work we assume the current SOBBH models from \citet{GWTC3_pop}, while for the population synthesis we employ the software package from \citet{Torrado_code}. Given the low frequency coverage of space interferometers, those detectors are sensitive to the inspiral phase of stellar-mass binaries, rather than to their mergers like LVK detectors. The population synthesis code generates catalogs of SOBBHs emitting in this frequency region according to the mass and spin distributions inferred from the third Gravitational-Wave Transient Catalog (GWTC-3), incorporating theoretical knowledge on star formation and evolution, such as the Madau-Fragos SFR \cite{Madau_2017}, as described in \citet{Torrado_SOBBH}. We assume the binary formation to be in a steady state, implying that the observed coalescence time is uniformly distributed across the population. In our analysis we set a maximum of 300 years for it, due to computational limitations. The redshift distribution is limited as well between $z=10^{-5}$ and $z=10$. As done for the MBHBs and the GBs, the orbit is assumed to be circular and we defer to Sec.~\ref{sec:caveats} for a general discussion of the impact of eccentricity.

\subsection{Extragalactic double white dwarfs}
\label{sec:extragalWD}

The visible universe hosts a very large population of stellar-mass binaries, which gives rise to a stochastic and isotropic GWB emitting in the milli-Hz window. The signal is primarily dominated by the DWD component up to $\approx \SI{40}{\milli\hertz}$ \cite{Staelens_Nelemans}. Unfortunately, extragalactic DWD systems may be completely impossible to resolve as individual sources, mainly due to their abundance and faint nature. \citet{Hofman_Nelemans} provided a fit for the shape of their GWB signal, in terms of the dimensionless energy density spectrum. We remark, however, that the amplitude and shape strongly depend on the astrophysical assumptions of the model \cite{Boileau_2025}. The level of this stochastic signal could exceed the instrumental sensitivity in the milli-Hz band, acting as a foreground signal that needs to be added to the instrumental noise.

%% file: methods.tex
\section{Methodology: global GWB estimate and SNR calculation}
\label{sec:methods}
We use the sensitivity curves and the synthetic catalogs described in the previous sections to estimate the unresolved GWB for the four detectors, through the simultaneous calculation of the GW signal and the signal-to-noise ratio (SNR) of all the systems present in the populations.

\subsection{Waveform}
We use appropriate waveform models to simulate the signal for each type of source. For MBHBs, we implement the phenomenological model PhenomC, which offers a valid approximation of the full inspiral-merger-ringdown signal for the quadrupole mode of coalescing BBHs with non-precessing (aligned) spins in a circular orbit \cite{Santamaria_PhC}. The waveform is computed in the frequency domain up to $10 f_{\text{ISCO}}$, where $f_{\text{ISCO}}$ is the orbital frequency at the Schwarzschild innermost circular orbit (ISCO),
\begin{equation}
    f_{\text{ISCO}} = \frac{1}{2\pi} \frac{c^3}{6\sqrt{6} G M} \; ,
\end{equation}
and $M$ is the total mass of the binary.

For GBs, SOBBHs and EMRIs, we instead take into account only the inspiral post-Newtonian signal, up to $f_{\text{ISCO}}$, as the coalescence phase does not contribute significantly to the GW emission measurable by the detectors under study for different reasons. For SOBBHs, the merger is at $f>10$ Hz, outside of the detectors' sensitivity band. Conversely, both EMRIs and GBs do merge in the frequency band of interest, but the merger-ringdown emission of the former is negligible due to the extreme mass ratio, while chances of observing a colliding WD-WD, WD-MS and MS-MS in our Galaxy within the detectors' lifetime is negligible.

We also remark that, except for EMRIs, we assume all sources to have zero eccentricity. For circular sources, we therefore evaluate only the dominant harmonic $n=2$. We will discuss our waveform choices further in Sec.~\ref{sec:caveats}.

\subsection{SNR calculation}
\label{sec:SNR_calculation}

We follow the formalism developed in \citet{Finn_Thorne} and the approach of \citet{Bonetti_EMRI}, especially relevant for eccentric binaries. For each class of binaries, the SNR is calculated as 
\begin{equation}
\label{eq:SNR}
    \mathrm{SNR}^2 = \int \frac{h_{c}^2}{f S_{n}(f)} \: d \ln f \; ,
\end{equation}
where $S_{ n}(f)$ is the detector PSD provided by Eq. ~\eqref{eq:PSD} with parameters given in Tab.~\ref{tab:noise} for each detector, and $h_c^2$ is the total characteristic strain of the signal, defined by the sum of the characteristic strain of all $n$ harmonics. For EMRIs, this is given by
\begin{equation}
\label{eq:hc_code_ecc}
    \begin{split}
        h_c^2(f) &= \frac{2}{3} \frac{G^{5/3}(\pi)^{2/3}M_{c}^{5/3}f^{-1/3}}{c^3 \pi^2 d^2} \times \Phi(f) \; ,\\
        \Phi(f) &= 2^{2/3} \sum_{n=1}^{\infty} \frac{g_n(e_n)}{n^{2/3}\mathcal{F}(e_n)} \; ,
    \end{split}
\end{equation}
where $\mathcal{F}(e_n)$ and $g_n(e_n)$ are analytical functions of the eccentricity (see \cite{Bonetti_EMRI} for details). We consider only the modes that corresponds to a rest-frame orbital frequency $f_{\text{orb}}= f(1+z)/n$, that lies in between the interval $\left[f_{\text{orb,start}}, f_{\text{orb,end}}\right]$. The minimum limit is the initial orbital frequency as given in the source catalogs, while the maximum limit is the final orbital frequency, which is either $f_{\text{ISCO}}$ if the EMRI plunges during the mission time or $f_{\text{orb}}(t=T_{\text{obs}})$ otherwise. The frequency evolution of eccentric binaries is computed using 
\begin{equation}
    \frac{df_{\text{orb}}}{dt} = \frac{96G^{5/3}}{5c^5}(2\pi)^{8/3}M_c^{5/3}f_{\text{orb}}^{11/3} \times \mathcal{F}(e) \; .
\end{equation}

Since MBHBs, GBs and SOBBHs are assumed on circular orbits, the computation of the characteristic strain is limited to the second harmonic $n=2$, becoming remarkably less demanding.

 As mentioned earlier, to properly include the coalescence phase, the amplitude of the signal for the MBHBs is provided by the PhenomC model up to $10 f_{\text{ISCO}}$, whereas the characteristic strain for GBs and SOBBHs is simply given by the inspiral post-Newtonian formula up to $f_{\text{ISCO}}$,
 
\begin{equation}
\label{eq:hc_code}
    h_c^2(f) = \frac{2}{3}\frac{G^{5/3}(\pi)^{2/3}M_{c}^{5/3}f^{-1/3}}{c^3 \pi^2 d^2} \; .
\end{equation}

Once an SNR value is assigned to each source, we can distinguish between resolved and unresolved sources by setting an SNR threshold for the individual detection. We use the standard SNR$=8$ threshold for all sources, except for EMRIs, for which we use a threshold SNR$=20$. This is because of the complex nature of their signals, which requires a higher SNR for confident detection, as demonstrated in mock data searches \cite{2010CQGra..27h4009B}.

One limitation of our SNR calculation is that it makes use of inclination-polarization averaged waveforms for the GW signals and sky-averaged PSD for the detectors. The advantage of these approximations is the firm gain in code performance, allowing us to analyze catalogs with millions of sources with minimal computational burden, while still providing a reasonable estimation of the unresolved GWB level. This has been demonstrated for EMRIs, for which a proper estimate of the GWB including the full detector response performed by \cite{Pozzoli_EMRI} has been found to be consistent with the much simpler estimate of \cite{Bonetti_EMRI} employed in this work.

\subsection{GWB computation}
\label{sec:GWB_computation}

Given the set of different catalogs of astrophysical sources, we estimate the total PSD of the stochastic background generated by all of them. The procedure closely follows the one explained in \cite{Bonetti_EMRI}, further simplified in the case of circular orbits.

In order to correctly assess the contribution of each single source to the overall signal, it is necessary to differentiate between sources that evolve or not in frequency during the observation time. In the second case, the characteristic amplitude of a GWB generated by eccentric binaries is
\begin{eqnarray}
\label{eq:hc_gwb2_ecc}
    &&h^2_{c,\text{gwb}}(f)= \int dz \; dM_c \; de \nonumber\\
    &&\times \left[ \sum_n \frac{d^4 N}{dz \; dM_c \; de \; d\ln{f_{\text{orb}}}} h^2_n(f) \right]_{f_{\text{orb}} = \frac{f(1+z)}{n}}  \; ,
\end{eqnarray}
where  $h_n $ is determined by the emitted power at that specific harmonic
\begin{equation}
    h_n = \sqrt{\frac{G \dot{E}_n}{c^3 \pi^2 d^2 f^2_n}} \; .
\end{equation}

In the case of circular binaries, Eq.~\eqref{eq:hc_gwb2_ecc} simplifies to 
\begin{equation}
\label{eq:hc_gwb2}
    h^2_{c,\text{gwb}}(f) = \int dz dM_c \; \frac{d^3 N}{dz \; dM_c \; d\ln{f_r}} h^2(f_r) \; ,
\end{equation}
where we are using the inclination-polarization averaged strain
\begin{equation}
\label{eq:inc_pol_averaged}
    h = \sqrt{\frac{32}{5}} \frac{(G M_c)^{5/3}}{c^4 r} (\pi f)^{2/3} \; .
\end{equation}

In both cases, $h^2_n$ and $h^2$ are evaluated in the frequency bin $\Delta f$ where the source is located, and the resulting contribution to $h_{c,\text{gwb}}$ is simply $h^2 \times f/\Delta f=h^2\times fT=h^2\times N_{\rm cyc}$, where $N_{\rm cyc}$ is the number of wave cycles completed within the observation time $T$.

If the frequency evolution of the binary spans a larger number of frequency bins during the observation time, we have to weight the contribution at each frequency with the ratio between the number of cycles ($f^2/\dot{f}$) covered by the source and the highest number of cycles observable ($fT_{\rm obs}$) at that frequency. This yields to these expressions for the characteristic strain of the GWB:
\begin{subequations}
\label{eq:whole}
\begin{equation}
\begin{split}
&h^2_{c,\text{gwb}}(f) = \frac{1}{2} \int dz \, dM_c \, de \\
& \times \left[ \sum_n 
\frac{d^4 N}{dz \, dM_c \, de \, d\ln f_{\text{orb}}} 
\frac{h^2_{c,n}(f)}{f T_{\text{obs}}} \right]_{f_{\text{orb}} = \frac{f(1+z)}{n}} \,,
\end{split}
\label{eq:hc2}
\end{equation}

\begin{equation}
h^2_{c,\text{gwb}}(f) = \frac{1}{2}\int dz \; dM_c \frac{d^3 N}{dz \; dM_c \; d\ln{f_{r}}} \frac{h^2_{c}(f_r)}{fT_{\text{obs}}}  \; ,
\label{subeq:1}
\end{equation}
\end{subequations}
where the first one applies to eccentric binaries, while the second one assumes zero eccentricity. The characteristic strain of the single source is determined using Eq. \eqref{eq:hc_code_ecc} and Eq. \eqref{eq:hc_code} for all the binary classes. 

As for the SNR calculation, the orbital frequency $f_{\text{orb}}$ has to be within the interval $\left[f_{\text{orb,start}},f_{\text{orb,end}}\right]$, which corresponds to limiting the harmonic numbers in the sum to the range $\left[n_{\text{min}},n_{\text{max}}\right]$,
\begin{equation}
        n_{\text{min}} = \frac{f(1+z)}{f_{\text{orb,start}}} \; , \quad
        n_{\text{max}} = \frac{f(1+z)}{f_{\text{orb,end}}} \; .
\end{equation}

We evaluate the detectability of the GWB signal by computing the SNR as  \cite{Sesana_2016}
\begin{equation}
\label{eq:SNR_gwb}
    \mathrm{SNR}_{\mathrm{gwb}}^2 = T_{\mathrm{obs}}\int \gamma(f) \frac{h_{c,\text{gwb}}^4}{f^2 S_{n}^2(f)} \: d f \; ,
\end{equation}
where $h_{c,\text{gwb}}$ is the GWB amplitude in characteristic strain, $S_{ n}(f)$ is the noise PSD of each detector, as defined in Eq. ~\eqref{eq:PSD} with parameters listed in Tab.~\ref{tab:noise}, and $\gamma(f)$ is approximated to unity. The output of Eq.~\eqref{eq:SNR_gwb} has to be considered only indicative for two reasons. First, the 
$\gamma(f)=1$ approximation holds only in the long wavelength approximation \citep{2013PhRvD..88l4032T} and, perhaps most importantly, for a single triangular constellation, the measurement relies on the possibility of performing a reliable independent noise estimate from the detector's null channel \citep{2019PhRvD.100j4055S}.

\subsection{Iterative procedure}
\label{sec:iterative_procedure}

One of the key features of our code is the capability of computing the detectable sources and the unresolved GWB of several astrophysical populations, taking into account the stochastic noise generated by all of them. This is achieved by implementing an iterative procedure that subtracts the brightest sources at each step.

During each repetition, the code begins by reading the four population catalogs (MBHBs, EMRIs, GBs and SOBBHs) and the sensitivity curve $S_n(f)$ of one of the four detectors (LISA, $\mu$Ares, DO, AMIGO), then it starts a cycle on the populations. 

At the first iteration $i=0$, the SNR value of each source in the catalog is computed as explained in Section \ref{sec:SNR_calculation}, considering the noise curve $S_n(f)$ given as input, which includes only the instrumental noise and the GWB from the extragalactic DWDs. These values of SNR are not relevant, as they do not take into account any noise from other GW sources. Then, the PSD $S_{h.0}(f)$ of the total GWB produced by all sources is calculated, following Section \ref{sec:GWB_computation} and stored in a file.
This process is looped on all populations, after which iteration $i=0$ is completed.

\begin{figure}[t]
    \centering
    \includegraphics[width=0.7\linewidth]{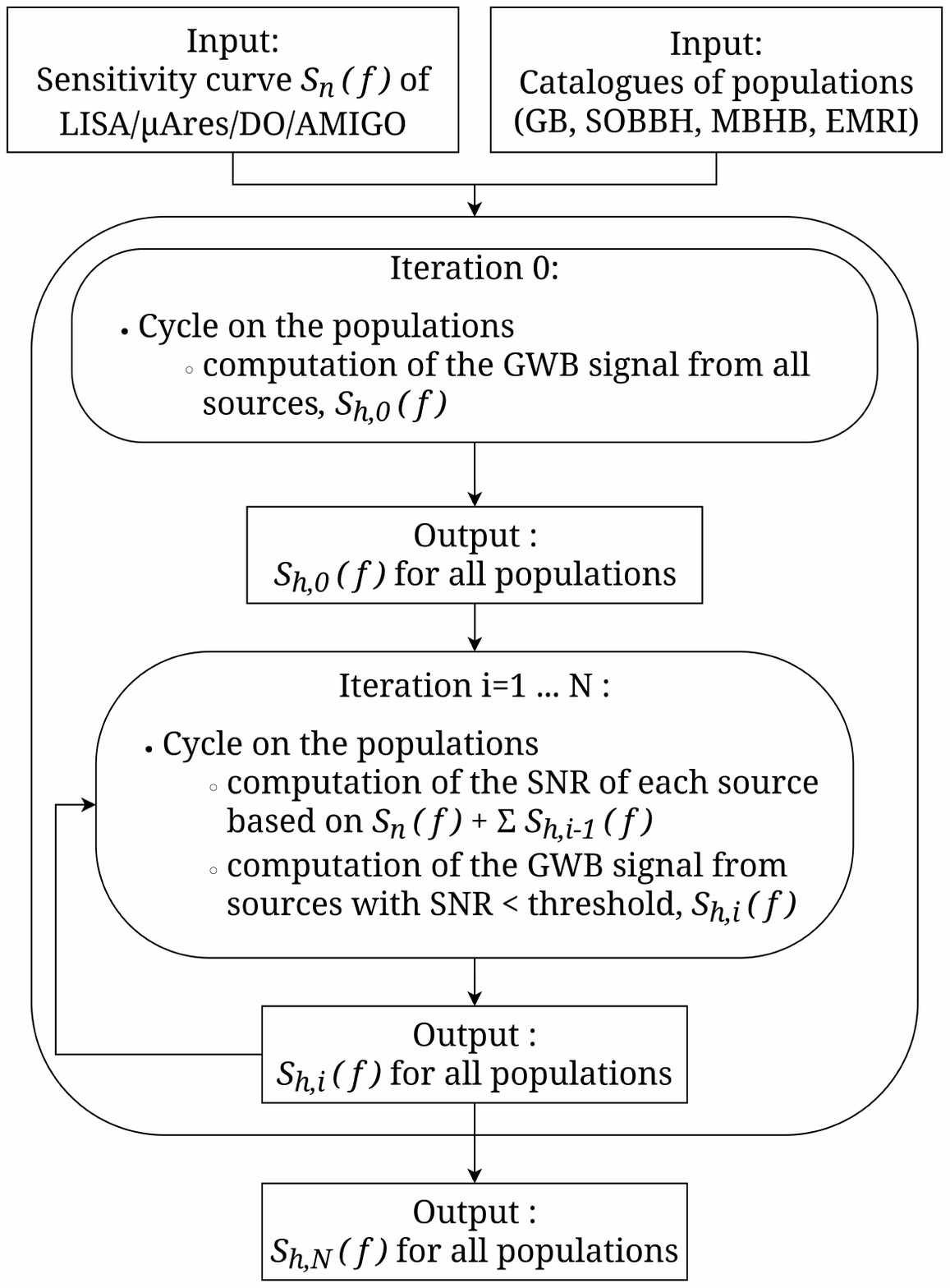}
    \caption{Flowchart of the code's algorithm.}
    \label{fig:flow_chart}
\end{figure}

In the next iteration $i=1$, the code reads the $S_{h,0}(f)$ arrays generated at step $i=0$  and it adds them to the detector sensitivity $S_n(f)$. Now the SNR of each GW event is determined with respect to the new noise curve $S_n(f) + \sum S_{h,0}(f) $ and is compared to the specified threshold for that catalog. The code then evaluates a new PSD of the GWB $S_{h,1}(f)$, by considering only the sources whose SNR is below the threshold, and prints it on file for each population. In each subsequent iteration $i$, the latest estimates $S_{h,i-1}$ is added to the detector noise and is used for the SNR calculation. A schematic overview of the code workflow is presented Fig.~\ref{fig:flow_chart}.

The algorithm repeats these steps until the number of detectable events in each catalog stabilizes to a constant value, meaning that our procedure has converged. Indeed, at each repetition, the level of the residual background reduces due to the higher number of resolved sources, since the SNRs progressively increase as a result of the lower noise given by the astrophysical background. We set a limit of $i=20$ iterations to reduce the runtime of the code: at this point, the number of resolved sources is no longer changing across all catalogs except for the GB population, where, in the worst case, it increases by only 0.2\% in the final iteration - a variation we consider negligible. Unlike other studies  \cite{Korol_WD, Karnesis_code}, our procedure does not apply a running median to smooth the PSD. We expect this to result in no significant difference on our final results, since applying a running median at the last step closely matches the non-smoothed curve. Higher attention on the smoothing may be required when fitting the foreground with an analytical model, which lies beyond the scope of this study.

At the final cycle $i=20$, the produced PSD $S_{h,20}$ is a more accurate estimate of the GWB produced by all the sources that are too faint to be identified and extracted by the collective signal. As a result, it provides a reliable prediction of the astrophysical noise that will impact the detector sensitivity. For this reason, we simply refer to it as $S_h$ in the following, and we use it to calculate the final SNR of the events in the catalogs.

%% file: results.tex
\section{Results}
\label{sec:results}
\subsection{Unresolved GW background}

One of the primary goals of this study is to evaluate the residual astrophysical background present in each detector, computed through the recursive subtraction outlined in Section~\ref{sec:iterative_procedure}. The five source classes that contribute to the stochastic signal are MBHB, EMRI, SOBBH, GB, and extragalactic DWD, with the latter included a priori in the iterative procedure, as an irreducible GWB. Since we consider two fiducial catalogs for the MBHB, derived from the different HS and LS  models, this results in two alternative outcomes for each mission design.

\begin{figure*}[t]
    \centering
    
    \begin{minipage}{0.49\textwidth}
        \centering
        \includegraphics[width=\linewidth]{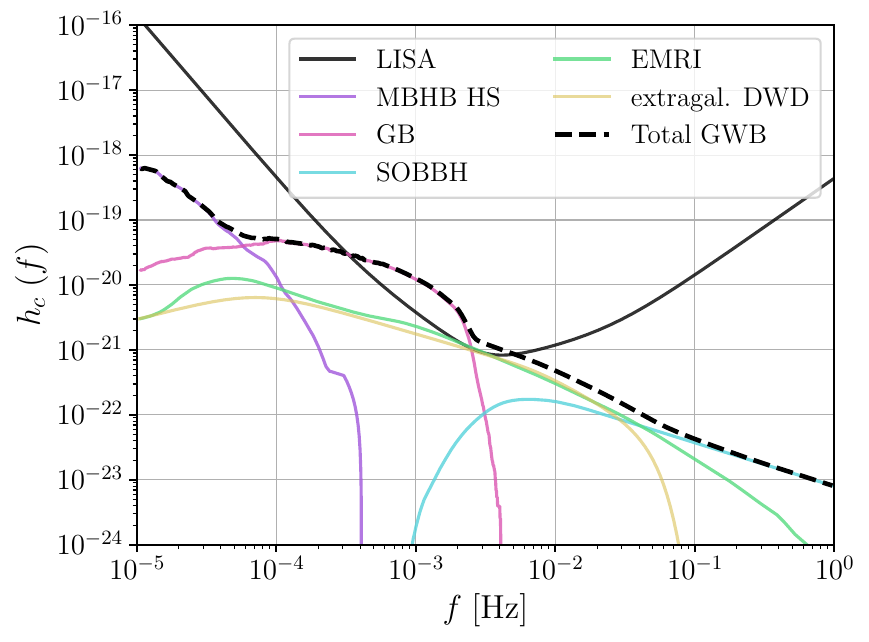}
    \end{minipage}
    \hfill
    \begin{minipage}{0.49\textwidth}
        \centering
        \includegraphics[width=\linewidth]{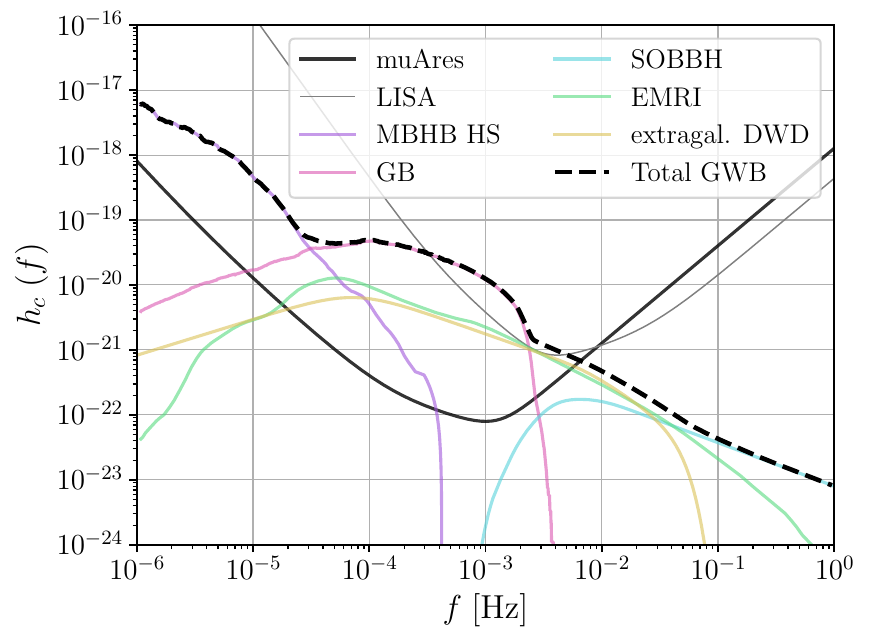}
    \end{minipage}
        \begin{minipage}{0.49\textwidth}
        \centering
        \includegraphics[width=\linewidth]{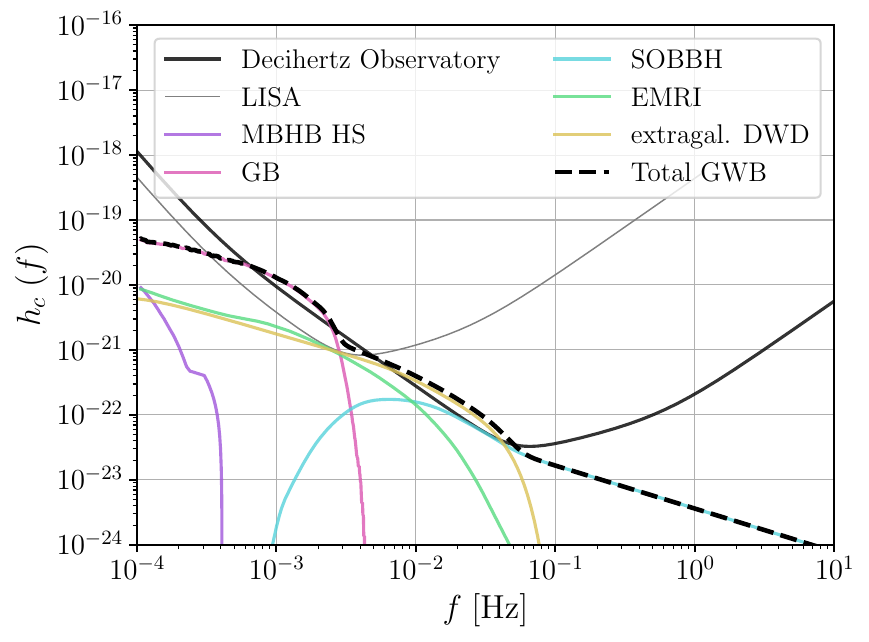}
    \end{minipage}
    \hfill
    \begin{minipage}{0.49\textwidth}
        \centering
        \includegraphics[width=\linewidth]{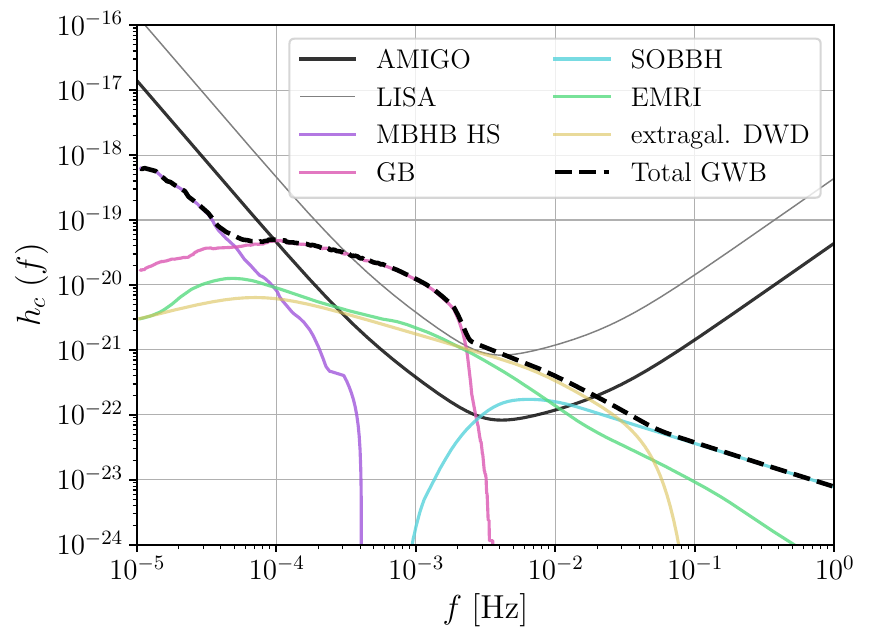}
    \end{minipage}

    \caption{Each panel shows the characteristic strain of the total unresolved GWB (black dashed line) present in the detector, computed in the HS scenario. The various contributions are represented by colored solid lines, while the solid black line denotes the instrumental PSD. The LISA noise curve is plotted in gray for comparison.}
    \label{fig:GWB_HS}
\end{figure*}

\begin{figure*}[t]
    \centering
    
        \begin{minipage}{0.49\textwidth}
        \centering
        \includegraphics[width=\linewidth]{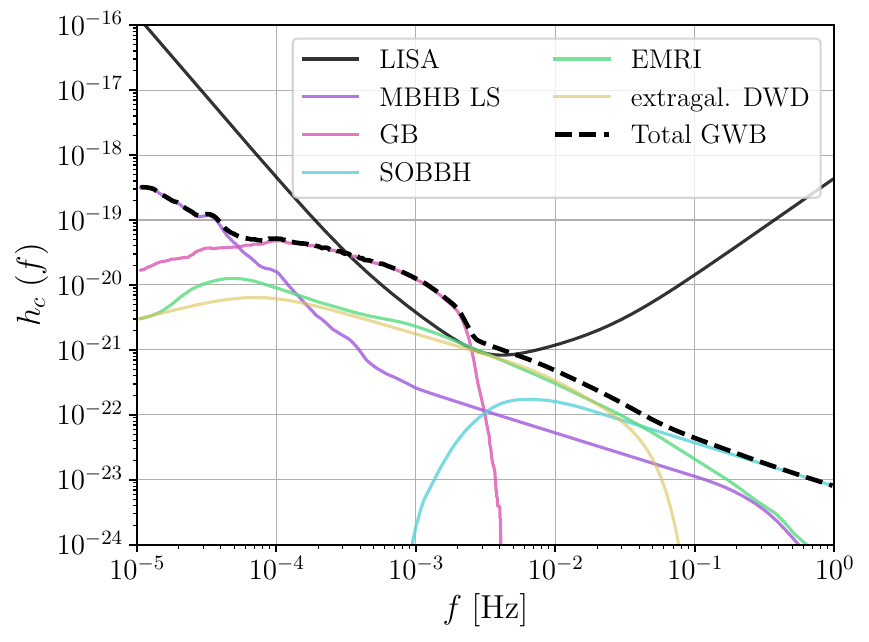}
    \end{minipage}
    \hfill
    \begin{minipage}{0.49\textwidth}
        \centering
        \includegraphics[width=\linewidth]{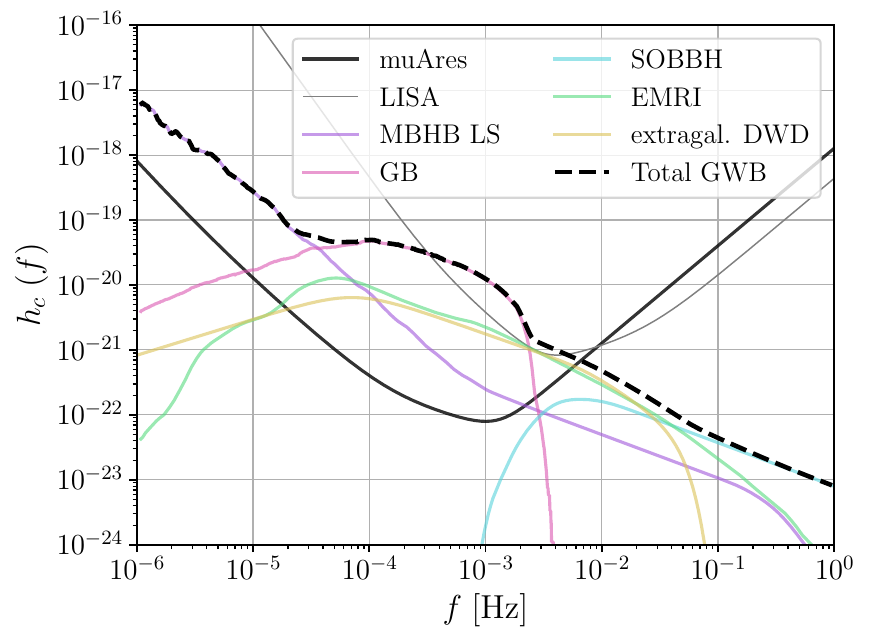}
    \end{minipage}
    \begin{minipage}{0.49\textwidth}
        \centering
        \includegraphics[width=\linewidth]{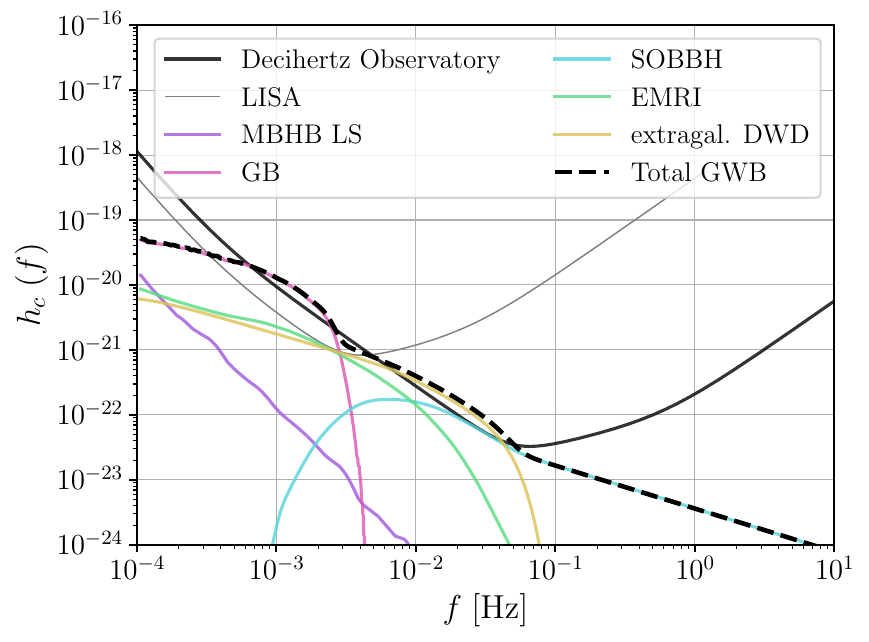}
    \end{minipage}
    \hfill
    \begin{minipage}{0.49\textwidth}
        \centering
        \includegraphics[width=\linewidth]{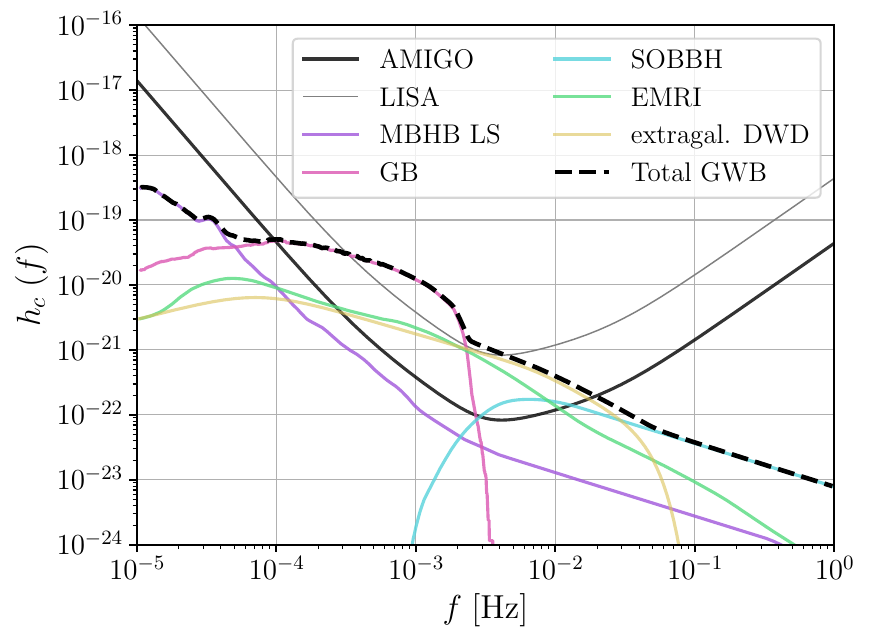}
    \end{minipage}

    \caption{Each panel shows the characteristic strain of the total unresolved GWB (dashed black line) present in the detector, computed in the LS scenario. The colored solid lines denote the contributions from different source populations. The instrumental PSD of each mission design is plotted in solid black, with the LISA one in gray for comparison.}
    \label{fig:GWB_LS}
\end{figure*}

In Fig.~\ref{fig:GWB_HS} (HS scenario) and Fig.~\ref{fig:GWB_LS} (LS scenario) we show the characteristic strain of the unresolved GWB $S_h(f)$ (colored lines) produced by the different populations at the last iteration of the code. A running mean has been applied in order to smooth the signal. Each plot displays also the analytic function for the stochastic background of extragalactic DWDs (yellow), detailed in Section~\ref{sec:extragalWD}, together with the cumulative sum of the five signals (dashed black), which represent our estimate of the total astrophysical GWB present in each detector. This signal must then be added to the instrumental noise curve (solid black) to obtain an estimate of the overall detector sensitivity.

At low frequency, the overall background is dominated by plenty of MBHBs far from coalescence, slowly inspiraling while emitting GW signals that are too faint to be individually resolved by the observatory. Their residual GWB signal overwhelms the instrumental PSD of $\mu$Ares, while it remains significantly below the ones of LISA, DO and AMIGO. It decreases to zero after $\sim \SI{10^{-4}}{\hertz}$ in the HS scenario, whereas it extends up to higher frequency in the LS case, due to the presence of less massive systems. In the milli-Hz region, the large number of compact binaries harboring in the Milky Way give rise to a Galactic foreground that exceeds the detectors' noise curves, before exhibiting an exponential decline due to the effective removal of the brightest sources emitting at high frequencies. DO is the least affected by this foreground, which has instead a great impact on LISA, $\mu$Ares and AMIGO. 

The stochastic background produced by a population of circular binaries driven by GW emission is characterized by a distinctive frequency dependence, scaling as $\propto f^{-2/3}$ (see e.g. \cite{Phinney}). In addition, it is generated by a superposition of feeble signals at low frequencies which gradually become brighter and less numerous at higher frequencies. Consequently, the deci-Hz region is least affected by astrophysical backgrounds, since the GWB level intrinsically decreases and individual signals can be more easily disentangled and resolved. The extragalactic DWD signal is the dominant contribution, before being surpassed by the SOBBH background. The former significantly exceeds the instrumental noise of AMIGO, while  only marginally approaching that of other missions. The EMRI GWB lies below or is comparable to the extragalactic DWD signal, depending on the detector’s efficiency in resolving sources. Along with the SOBBH GWB, it is strongly reduced by DO, due to the large number of extracted signals.

For each detector, we compute the SNR of the unresolved GWB of each source population, as well as of the total residual GWB, using Eq. \eqref{eq:SNR_gwb}. For the former calculations, the considered noise PSD includes the instrumental sensitivity and the sum of the unresolved GWBs of the other populations. In the latter case, the noise PSD only includes the instrumental sensitivity. The observation time is set to 10 years. Tab.~\ref{tab:snr_values} summarizes the results. Each contribution is individually detectable by the different missions, but for the MBHB population. The latter is detectable by $\mu$Ares with an SNR of the order of tens of thousands for both the HS and LS cases. We remark that these SNR values must be considered only as a \textit{theoretical} estimate of the detectability of the GWB signal, applicable when the instrumental noise is perfectly known or when two detectors are available to perform cross-correlation. Indeed, as noted in Sec.~\ref{sec:GWB_computation}, Eq.~\ref{eq:SNR_gwb} ignores the difficulty of distinguishing between the instrumental noise and the GWB itself.

Astrophysical GWBs are themselves interesting signals, but they act as an irreducible noise level for the identification of individual deterministic sources as well as for the perhaps more interesting cosmological backgrounds. With this in mind, the numbers in Tab.~\ref{tab:snr_values}, together with the GWB visualization of Figs.~\ref{fig:GWB_HS} and~\ref{fig:GWB_LS}, provide useful guidance for sizing future space missions that efficiently maximize scientific return. In this respect, LISA and DO are well suited to their respective frequency bands, yielding the lowest SNR of the residual astrophysical GWB. In particular, DO appears to be nearly optimal as it "touches" the irreducible galactic and extragalactic DWD GWB across two decades in frequency, from $10^{-3}$Hz to 0.1 Hz. AMIGO, despite the tenfold increase in sensitivity with respect to LISA, is unlikely to outperform LISA in the $10^{-4}-10^{-2}$Hz range, because it will quickly run into the DWD foregrounds. Finally, although $\mu$ARES will clearly outperform LISA below $10^{-4}$ Hz, the MBHB GWB, regardless of the seeding model, will limit its sensitivity by more than one order of magnitude at those low frequencies. This means that mission requirements can be optimized to make it more feasible with minimal scientific loss in the micro-Hz band compared to the original mission design.

\begin{table*}[t]

\begin{ruledtabular}
\begin{tabular}{c|c|c|c|c|c|c|c}
\textrm{ }&
MBHB HS&
MBHB LS&
EMRI&
SOBBH&
GB&
extragalactic DWD&
Total\\

\colrule
LISA & 0.13 & 6.5 & 450 / 447 & 32 & 3814 / 3802 & 458 / 455 & 4121 / 8134 \\
$\mu$Ares & 40964 & 13470 & 765 / 757 & 40 & 14974 / 14567  & 911 / 897& $2.36\times10^7$ \\
DO & 0.01 & 0.1 & 248 & 2092 & 1239 & 3288 & 9518\\
AMIGO & 3.6 & 4.6 & 584 & 462 & 13087 / 12956 & 3853 / 3840 & $7.33\times10^5$ \\
\end{tabular}
\end{ruledtabular}
\caption{\label{tab:snr_values}
SNR values of the unresolved GWB of the different source populations and of the total GWB for each detector, assuming a 10-year observation. We report the results for the HS and LS scenarios, respectively, when they differ.}
\end{table*}

\begin{table*}[t]
    \centering
    \begin{tabular}{c|c|c|c|c|c}
        \midrule
        \midrule
        & MBHB HS & MBHB LS & EMRI & SOBBH & GB \\
        \hline
        LISA & 226 & 563 & 567 & 2 &  20787 \\
        $\mu$Ares & 473 & 798 &  912 & 2 & 23815 \\
        DO & 226 & 2140 & 6967 & $2.95\times10^5$ & 16094\\
        AMIGO & 237 & 1747 & 5795 & 756 & 24402\\
        \midrule
        \midrule
    \end{tabular}
    \caption{ Number of resolved sources in the different populations, for each detector, assuming a 10-year observation.}
    \label{tab:resolved_sources}
    
\end{table*}

\subsection{Resolved sources}
\label{sec:resolved}

The overall GWB discussed in the previous section is used as an estimate of the astrophysical foreground noise in the final SNR calculation. The code delivers the predicted SNR value for each event in the catalogs, allowing us to distinguish between detectable and undetectable systems. For the first category, we analyze their parameter distributions in order to infer possible defining properties. The exact numbers of resolved sources are reported in Tab.~\ref{tab:resolved_sources}. We recall that we choose the following values of the SNR threshold: 20 for EMRIs and 8 for MBHBs, GBs and SOBBHs.

\subsubsection{Massive black hole binaries}
\begin{figure}[b]
    \centering
    \includegraphics[width=\linewidth]{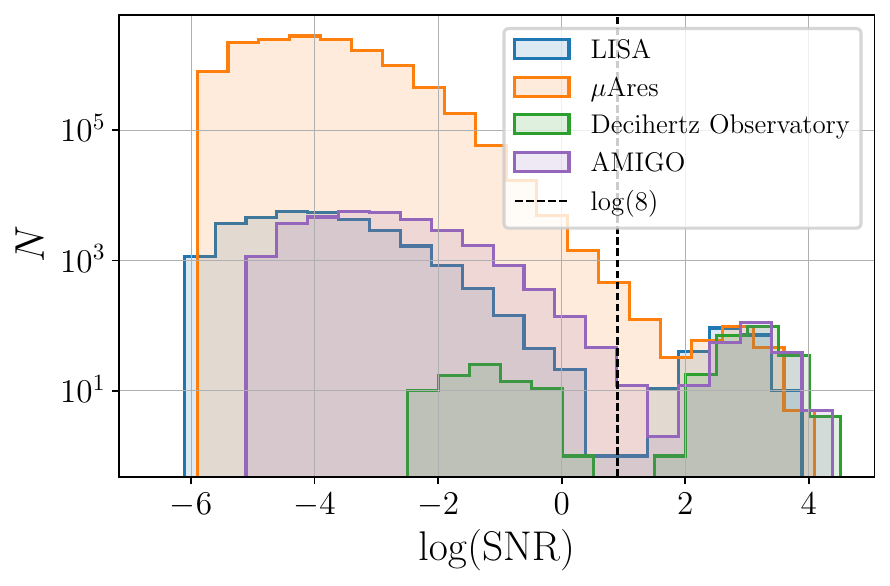}
    \caption{ SNR distribution for the MBHB catalog in the HS scenario.}
    \label{fig:SNR_MBHB_HS}
\end{figure}

\paragraph*{Heavy seeds.}
We first concentrate on the HS catalog, which contains $\approx1.4 \times 10^7$ sources. The resulting SNR distributions for the four missions are shown in Fig.~\ref{fig:SNR_MBHB_HS}. The different heights of the histograms come from the fact that for each detector we considered only MBHBs emitting at frequencies falling within its frequency band, setting the SNR of all other systems to zero.

\begin{figure*}[t]
    \centering

    \includegraphics[width=\linewidth]{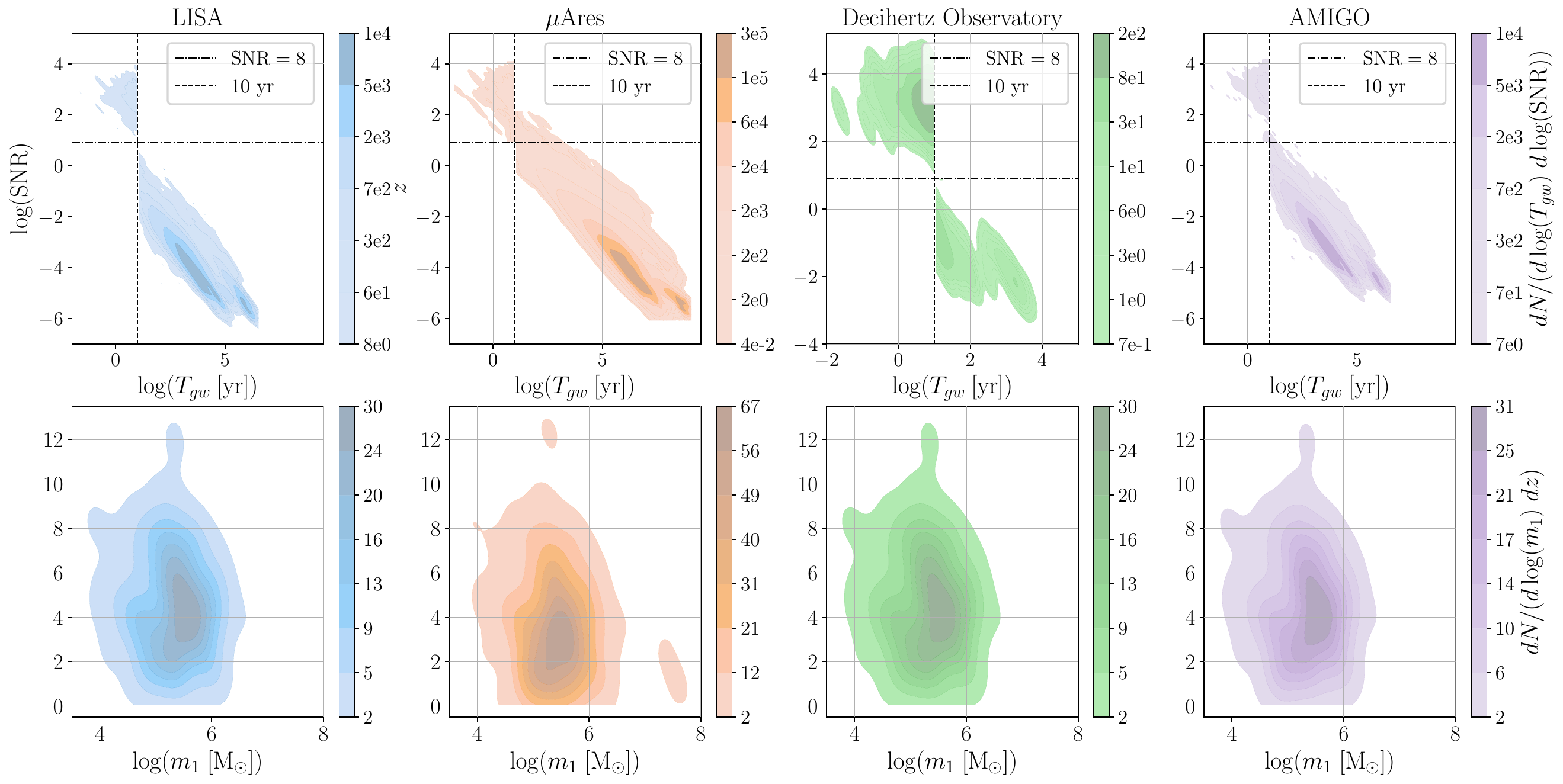}
    \caption{Upper panels: coalescence time (\textit{x}-axis) and SNR (\textit{y}-axis) distributions for the MBHB catalog in the HS scenario. All sources below the horizontal line marking SNR$=8$ are not resolved and contribute to the GWB. Lower panels: rest-frame primary mass (\textit{x}-axis) and redshift (\textit{y}-
axis) distributions for the resolved MBHBs in the HS scenario.}
    \label{fig:Tgw_SNR_mass_z_MBHB_HS}
\end{figure*}

Both the distributions for LISA and DO exhibit a clear cut corresponding to the SNR threshold, strictly separating between resolved and unresolved sources. This boundary is strongly related to the coalescence time $T_{gw}$ of the binary, as demonstrated by the upper panels of Fig.~\ref{fig:Tgw_SNR_mass_z_MBHB_HS}. In the case of LISA and DO, almost all the events with SNR $>8$ have $T_{gw}<\SI{10}{\year} $ and vice versa, implying that these detectors are capable of resolving only the binaries that reach the merger phase within the observation time, which correspond to slightly more than 200 systems in the HS catalog. AMIGO has similar capabilities, resolving only $\approx 10$ non-merging events.

On the other hand, $\mu$Ares is able to resolve $\approx 250$ additional non-merging sources, as we can see in Fig.~\ref{fig:Tgw_SNR_mass_z_MBHB_HS}. Thanks to its sensitivity curve extending to lower frequencies, it accumulates enough SNR already in the adiabatic inspiral phase, making it possible to detect MBHBs that are very distant from their coalescence, up to $\gtrsim\SI{10^2}{\year} $ before. 

\begin{figure}[t]
    \centering
    \includegraphics[width=\linewidth]{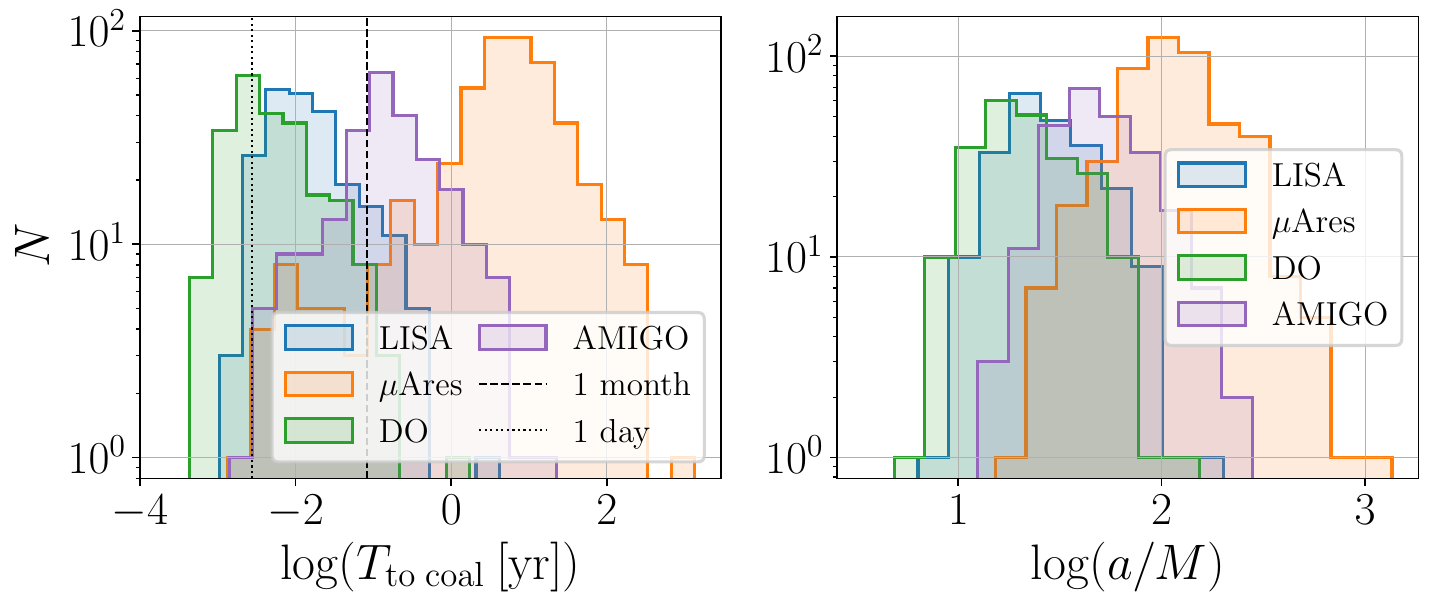}
    \caption{ Time to coalescence and binary separation distributions at the moment of detection for the MBHB catalog in the HS scenario.}
    \label{fig:Tdet_MBHB_HS}
\end{figure}

We explore the mass and redshift distributions of the resolved events in the lower panels of Fig.~\ref{fig:Tgw_SNR_mass_z_MBHB_HS}. The merging binaries detected by LISA, DO and AMIGO, have primary masses $m_1$ mainly confined to the interval between $\sim \SI{10^4}{\solarmass}$ and $\sim \SI{10^6}{\solarmass}$, while the furthest detectable sources extend to $z \approx 12$. The additional binaries identified by $\mu$Ares are primarily found at low redshift and tend to be more massive, exceeding $\SI{10^7}{\solarmass}$. 

Finally, it is interesting to further investigate the properties of the detectable systems at the time their signal can be identified in the data, which is when they reach the threshold SNR=8.  As shown by Fig.~\ref{fig:Tdet_MBHB_HS}, LISA and DO start to spot the binaries from a few weeks to days prior to the merger, whereas AMIGO detects them roughly a few months in advance. On the contrary, $\mu$Ares has the capability of detecting merging sources significantly earlier compared to the other detectors. This is reflected in the distributions of the binary separation $a/M$ (i.e. in units of the binary gravitational radius) at the moment of nominal detection (i.e. when SNR=8). Figure~\ref{fig:Tdet_MBHB_HS} shows how the histogram for $\mu$Ares is shifted approximately by one order of magnitude to the right, indicating that it resolves MBHBs at an earlier stage of their evolution, probing separations of $a/M$ on the order of a few hundred. AMIGO also shows an improvement over LISA and DO, detecting binaries earlier in their evolution.

\begin{figure}[t]
    \centering
    \includegraphics[width=\linewidth]{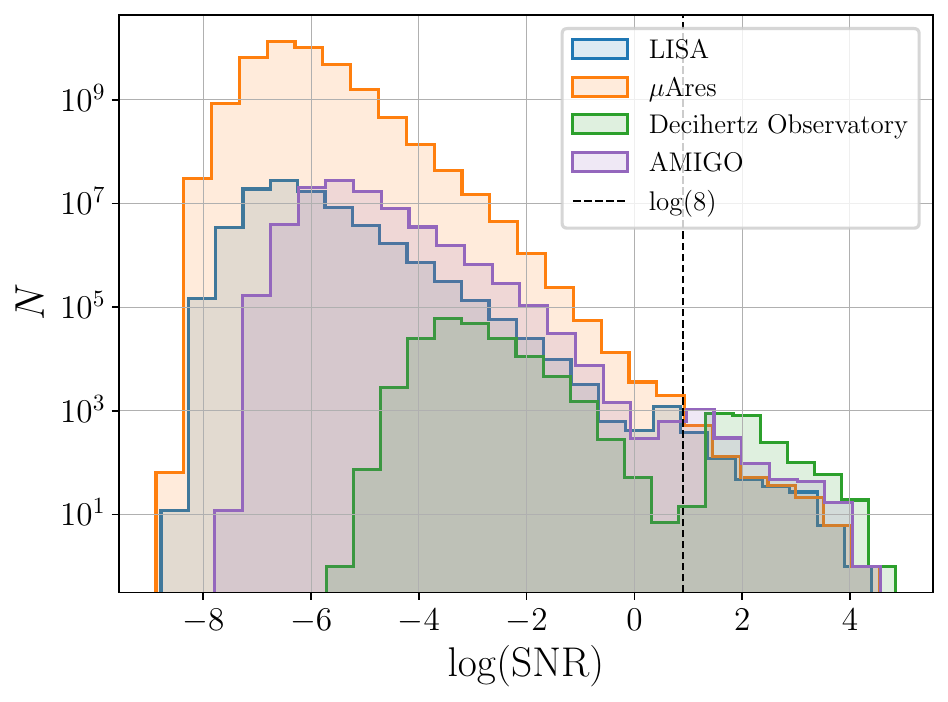}
    \caption{  SNR distribution for the MBHB catalog in the LS scenario.}
    \label{fig:SNR_MBHB_LS}
\end{figure}

\begin{figure*}[t]
    \centering
    \includegraphics[width=1\linewidth]{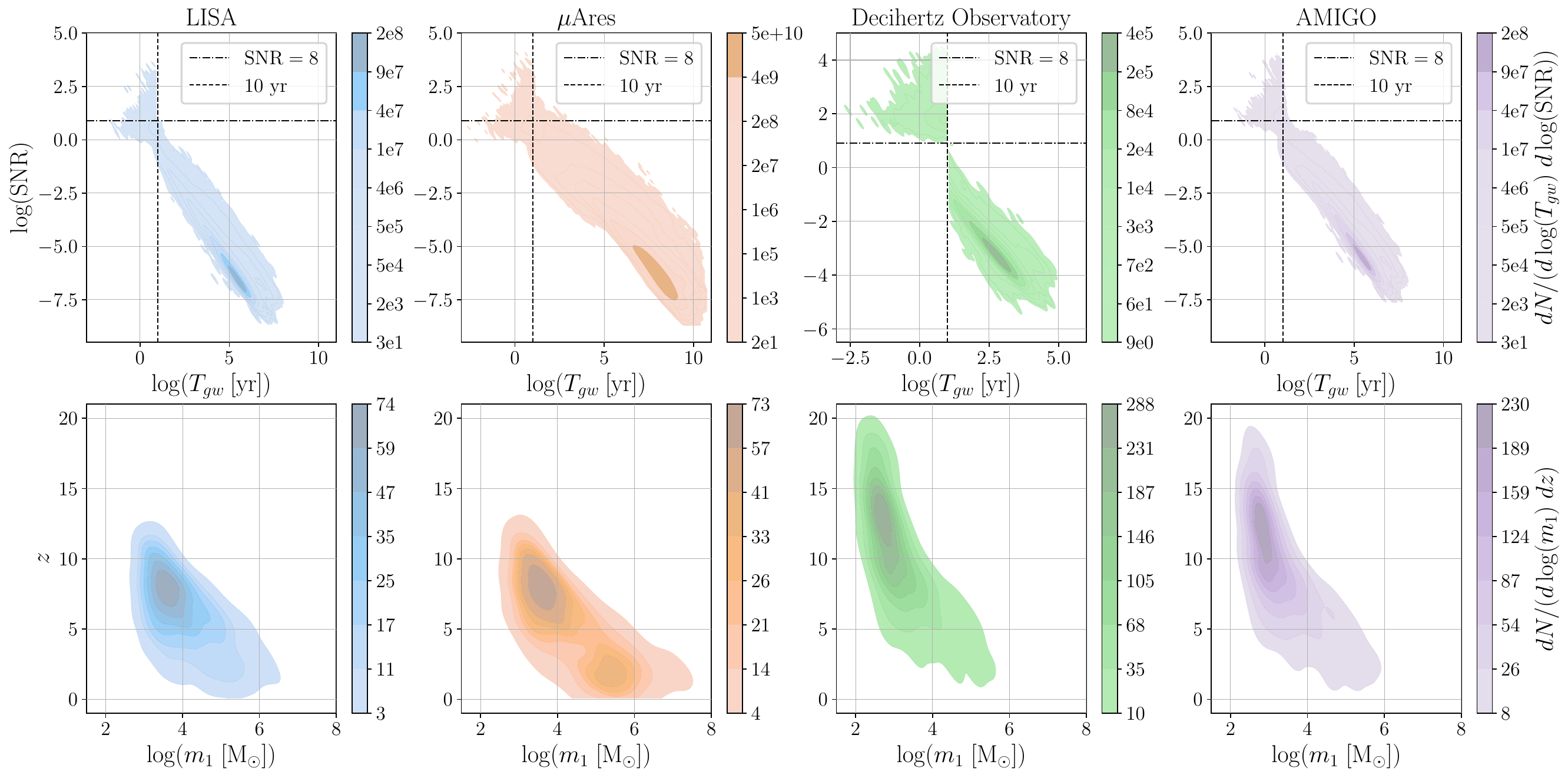}
    \caption{ Upper panels: coalescence time (\textit{x}-axis) and SNR (\textit{y}-axis) distributions for the MBHB catalog in the LS scenario. Lower panels: rest-frame primary mass (\textit{x}-axis) and redshift (\textit{y}-axis) distributions for the resolved MBHBs in the LS scenario.}
    \label{fig:Tgw_SNR_mass_z_MBHB_LS}
\end{figure*}

\paragraph*{Light seeds}
Now we turn our focus to the LS model, which assumes the existence of binaries composed of lighter MBHs of a few hundreds of solar masses. The total number of binaries is also much higher, consisting in around 40 billions of sources with an orbital frequency that spans from $\SI{10^{-6}}{\hertz}$ to $\SI{10^{-1}}{\hertz}$. Figure~\ref{fig:SNR_MBHB_LS} presents the overall SNR distribution for the four missions, while the correlation between SNR and $T_{gw}$ is shown in Fig.~\ref{fig:Tgw_SNR_mass_z_MBHB_LS}.

We can observe different situations for different detectors. Starting with LISA, the clear separation between merging and non-merging sources is no longer observed. This is because not all the sources reaching the coalescence phase are detectable by LISA. Some of them have such low mass or high redshift values that their signal does not reach the SNR threshold. LISA identifies roughly a quarter of the $\approx 2000$ merging binaries. The same is true for $\mu$Ares, which misses quite the same amount of merging sources, but detects $\approx200$ slowly chirping MBHBs, as in the HS case. In contrast, DO successfully detects all merging systems, while missing the inspiraling ones; therefore, it shows the typical bimodal SNR distribution also seen for the HS model. 
Finally, AMIGO sits somewhat in between $\mu$Ares and DO, by resolving $\sim 80\%$ of the merging binaries, together with a dozen of non-merging ones.

\begin{figure}[t]
    \centering
    \includegraphics[width=1\linewidth]{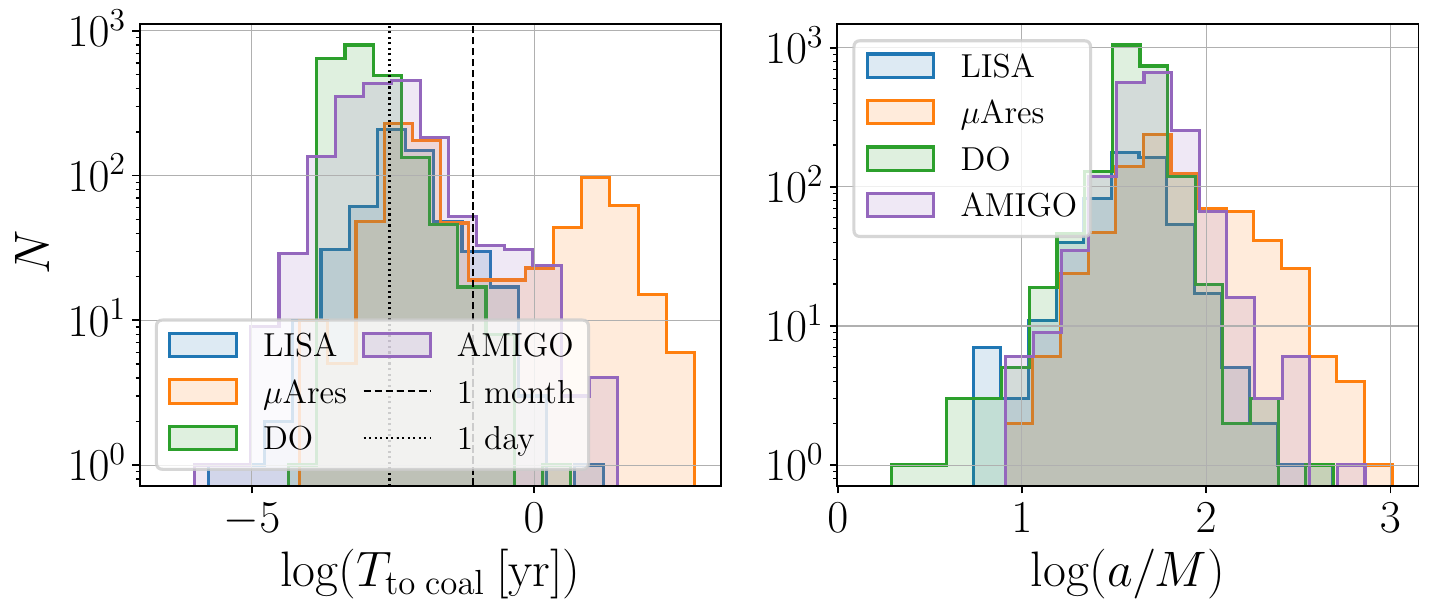}
    \caption{ Binary separation distributions at the moment of detection for the MBHB catalog in the LS scenario.}
    \label{fig:Tdet_MBHB_LS}
\end{figure}

The various classes of sources observed by each mission are clearly visible in the lower panels of Fig.~\ref{fig:Tgw_SNR_mass_z_MBHB_LS}. The LS model predicts the presence of lighter MBH newly formed in the early universe, which accrete mass throughout their evolution, ultimately resulting in more massive systems located at low redshift. LISA manages to identify merging MBHs with mass of the order of $10^3 - \SI{10^5}{\solarmass}$ up to redshift 12 and a few heavier systems within $z \approx 5$. For the latter, $\mu$Ares exceeds LISA observations by resolving a plenty of inspiraling MBHs with masses above $\SI{10^5}{\solarmass}$, mainly inhabiting the nearby universe. Instead, the strength of DO and AMIGO is the detection of numerous MBH light-seeds, freshly-formed remnants of population-III stars, at redshift above 15, along with a large number of merging MBHs between $z \approx 10$ and $z \approx 15$ with masses below $\SI{10^3}{\solarmass}$.

\begin{figure}[t]
    \centering
    \includegraphics[width=1\linewidth]{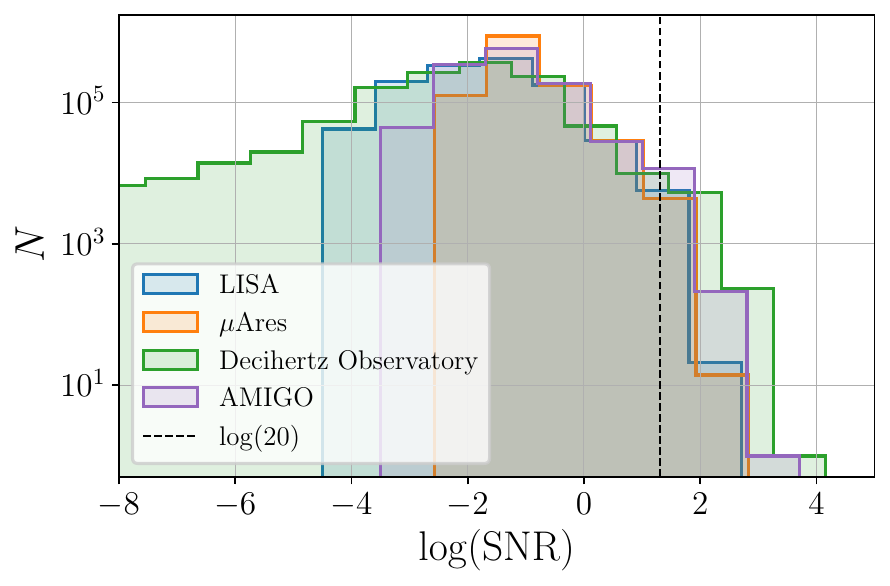}
    \caption{ SNR distribution for the EMRI catalog.}
    \label{fig:SNR_EMRI}
\end{figure}

\begin{figure*}[t]
    \centering
    \includegraphics[width=1\linewidth]{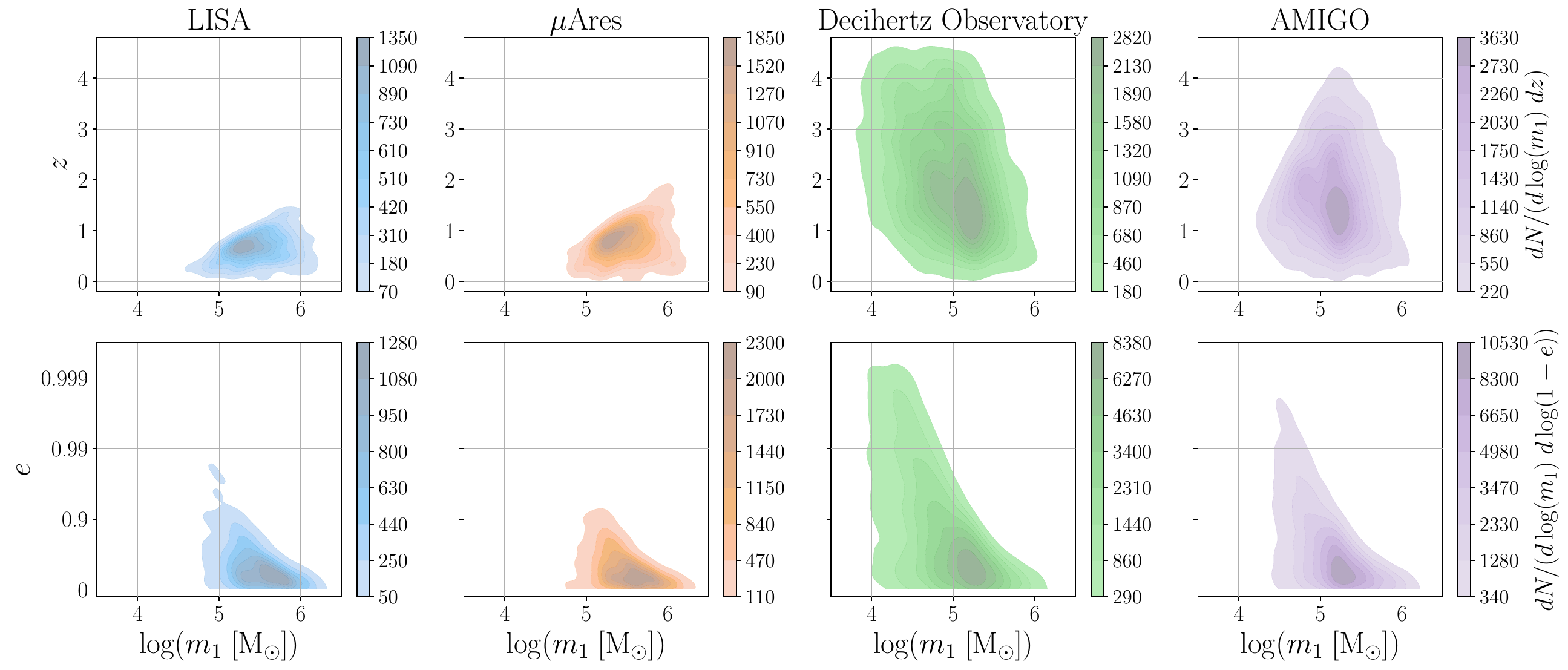}
    \caption{ Mass (\textit{x}-axis), redshift (\textit{y}-axis, upper panels) and eccentricity (\textit{y}-axis, lower panels) distributions for the resolved EMRI events.}
    \label{fig:mass_z_e_EMRI}
\end{figure*}

As for the HS case, we examine the binary separation at which the GW signal starts to be measurable by the instrument. From Fig.~\ref{fig:Tdet_MBHB_LS} we observe that the distributions for the different missions show greater overlap compared to the previous scenario. DO is now sensitive to the lighter systems and is able to detect them more in advance in their evolution. As in the previous case, $\mu$Ares begins to resolve the more massive sources in their very early inspiral, at separations of $a/M > 100$.

\subsubsection{Extreme mass-ratio inspirals}

Mock populations of EMRIs are difficult to construct due to our limited knowledge on the number and properties of these systems. Therefore our results are closely tied to the fiducial model we are considering, which contains a total of $1.2 \times 10^6$ sources. Approximately 550 of them are identified by LISA, while $\mu$Ares detects about 350 more events. DO and AMIGO resolve nearly 7000 and 6000 systems respectively, as shown by the SNR distribution in Fig.~\ref{fig:SNR_EMRI}.

Most EMRIs resolved by LISA and $\mu$Ares orbit a central MBH of mass $m_{1}$ between $\SI{10^5}{\solarmass}$ and $\SI{10^6}{\solarmass}$, while their eccentricity $e$ is mostly limited to $\log(1-e) \gtrsim -1$. Figure~\ref{fig:mass_z_e_EMRI} clearly demonstrates the capability of DO and AMIGO to resolve many more sources, characterized by lower mass values, down to $\sim \SI{10^4}{\solarmass}$ for DO, and notably higher eccentricities, up to $\log(1-e) \approx -3$. We must point out that most eccentric binaries are intentionally ignored in our analysis due to their expensive computational cost, since their GW signal involves a great number of harmonics. Still, the total amount corresponds to less than $1\%$ of the total population; thus, we can safely neglect those EMRIs for the goal of this study. Nevertheless, it is true that the eccentricity distribution of the resolved sources may extend to larger values, up to the limit $\log(1-e) \approx -5$ of the population, a feature that might be taken into account for the development of waveforms. Finally, DO is also capable of resolving EMRIs very distant in redshift, unlike LISA and $\mu$Ares, which stop at $z \approx 2$. AMIGO is able to arrive at $z \approx 4$ for the most massive systems. The limit $z=4.5$ in Fig.~\ref{fig:mass_z_e_EMRI} is constrained by the EMRI catalog we are employing, as discussed in Section~\ref{sec:sources}. We note, however, that by that redshift, the number density of sources has decreased by more than one order of magnitude compared to the peak at $z\gtrsim 1$, meaning that we are likely missing only a few percentage of the total count of detectable EMRIs.

\subsubsection{Stellar-origin binary black holes}

\begin{figure}[t]
    \centering
    \includegraphics[width=1\linewidth]{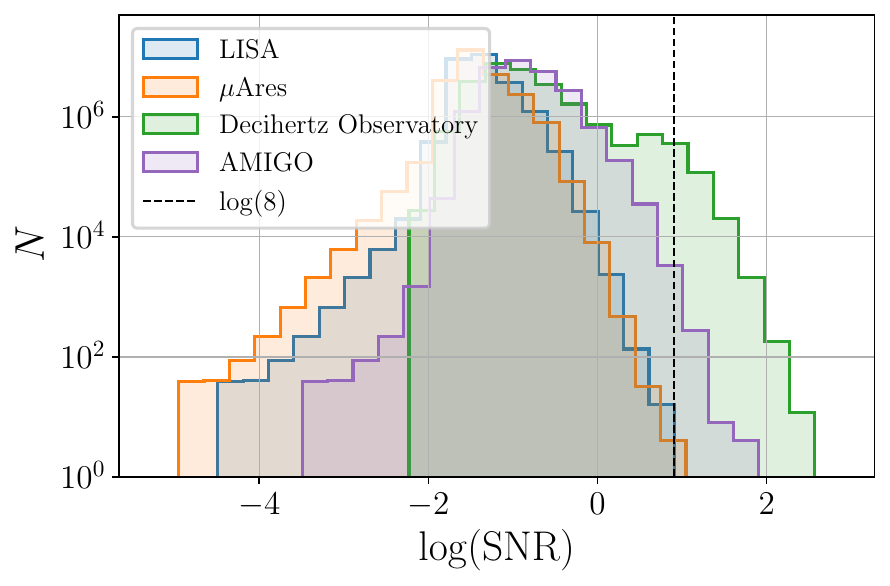}
    \caption{ SNR distribution for the SOBBH catalog.}
    \label{fig:SNR_SOBBH}
\end{figure}

Compact binaries of stellar BHs are already the primary target of terrestrial observatories, but they could be potentially detected also by space-based interferometers. In our model, they amount to nearly 26 million of events included in the catalog, whose SNR histograms is displayed in Fig.~\ref{fig:SNR_SOBBH}. As expected, almost all the SNR values for LISA and $\mu$Ares are below the threshold, with only a couple of events loud enough to be measured. In contrast, AMIGO yields about $\approx 750$ detections, while DO can  resolve all merging systems up to $z\gtrsim10$, amassing $\approx 3\times10^5$ detections.

\begin{figure}[t]
    \centering
    \includegraphics[width=1\linewidth]{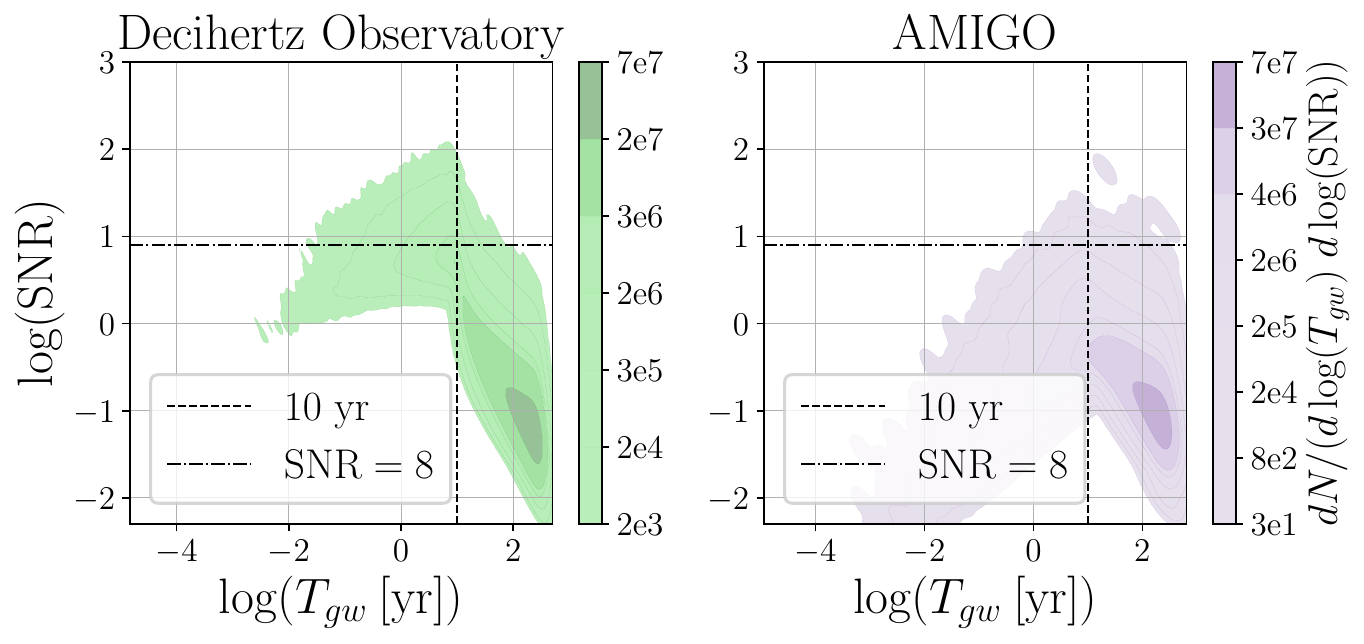}
    \caption{ Coalescence time (\textit{x}-axis) and SNR (\textit{y}-axis) distributions of the SOBBH catalog for DO and AMIGO.}
    \label{fig:Tgw_SNR_SOBBH}
\end{figure}

In Fig.~\ref{fig:Tgw_SNR_SOBBH} we plot the SNR values as a function of the coalescence time. For DO, we recognize a similar pattern as the one for MBHBs, where the majority of the resolved sources are binaries that merge during the observation time. These systems indeed span a large array of frequencies during their evolution, crossing the detector sensitivity bucket, although their coalescence phase is not measured by DO. However this condition is not always met, as there are binaries that are extremely close to the merger, already at too high frequencies to be detectable, as well as a fraction of non-merging systems that can be resolved. On the other hand, the systems detected by AMIGO are binaries with coalescence times comparable to the mission duration. Their signals remain in band throughout the observation, while still evolving across a sufficiently broad range of frequencies.

\begin{figure}[t]
    \centering
    \includegraphics[width=1\linewidth]{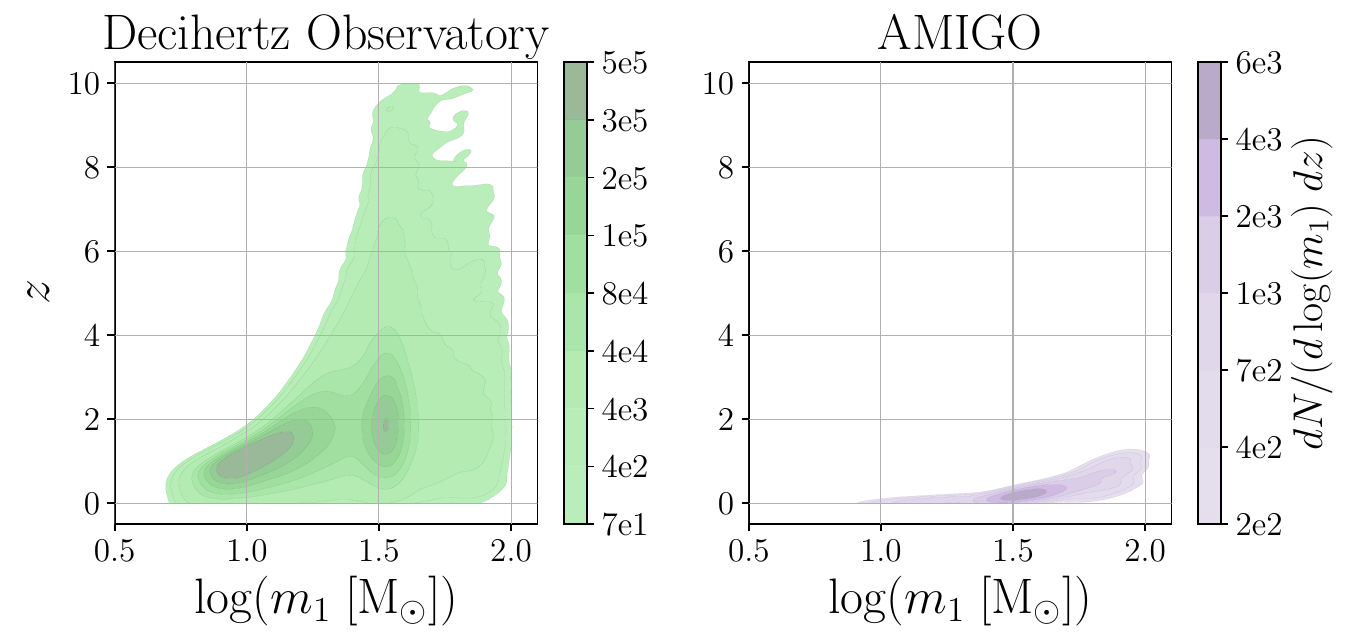}
    \caption{ Mass (\textit{x}-axis) and redshift (\textit{y}-axis) distributions of the resolved SOBBH events for DO and AMIGO.}
    \label{fig:mass_z_SOBBH}
\end{figure}

The primary mass and redshift values of the resolved sources for DO and AMIGO are displayed in Fig.~\ref{fig:mass_z_SOBBH}. For the most massive systems, the redshift horizon of DO extends up to $z\approx10$, which corresponds to the upper limit of the population considered here, whereas for AMIGO it remains below $z \approx 1$. The few binaries detected by LISA and $\mu$Ares are in the local universe, at $z\approx0$. 

\subsubsection{Galactic binaries}

\begin{figure}[t]
    \centering
    \includegraphics[width=1\linewidth]{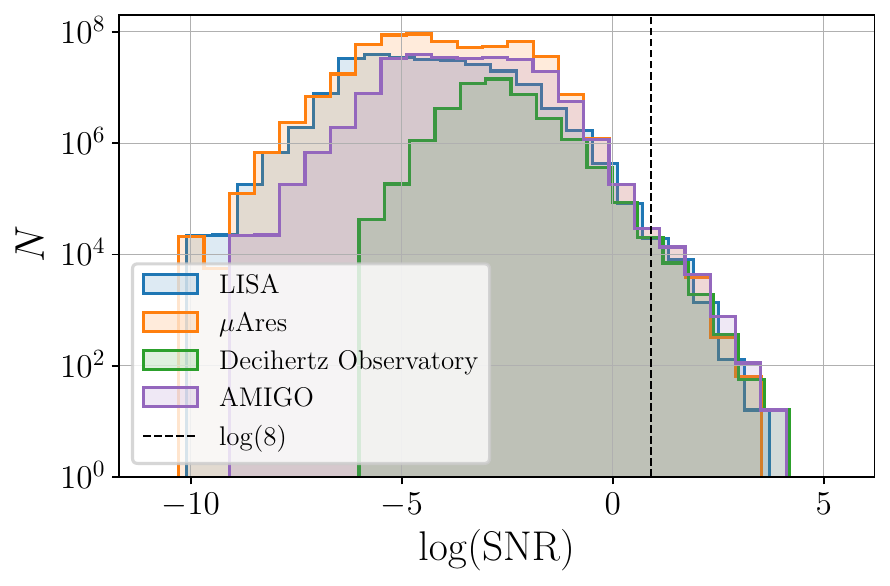}
    \caption{ SNR distribution for the GB catalog.}
    \label{fig:SNR_GB}
\end{figure}

\begin{figure}[t]
    \centering
    \includegraphics[width=1\linewidth]{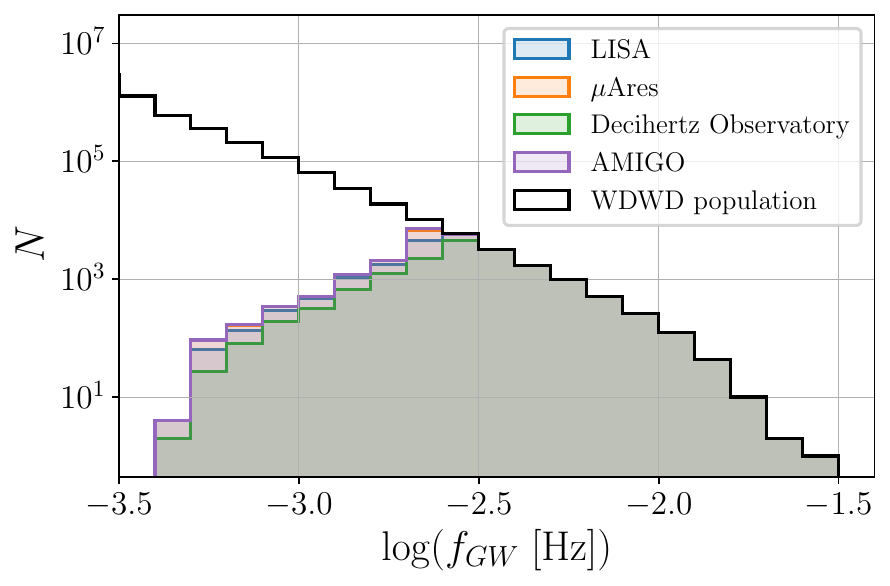}
    \caption{GW frequency distribution for the resolved WDWD systems, compared to the distribution of the whole WDWD population (black line).}
    \label{fig:fgw_GB}
\end{figure}

Our last astrophysical population consists of around 500 million of compact binaries inhabiting the Milky Way, whose quasi-monochromatic GW signals span a frequency range from $\SI{10^{-6}}{\hertz}$ to a few tens of $\SI{}{\milli \hertz}$. The SNR distribution of the "in-band" sources for each detector is presented in Fig.~\ref{fig:SNR_GB}. All resolved sources are WDWD binaries, whereas WDMS and MSMS systems contribute only to the unresolved GWB. The performance of the four detectors is rather similar, each resolving $O(10^4)$ systems over a 10-year observation time: AMIGO and $\mu$Ares detect the maximum number of sources ($\approx 24000$), followed by LISA ($\approx 21000$) and DO ($\approx 16000$). The larger number of sources detected by the former two is concentrated at lower frequencies, while each mission provides a complete survey of the population at frequency above $\SI{3}{\milli \hertz}$, as shown in Fig.~\ref{fig:fgw_GB}. No significant differences are found in the other properties of the resolved sources between the four detectors.

%% file: discussion_conclusions.tex
\section{Discussion}
\label{sec:discussion_conclusions}

\subsection{GWB across bands}
In this study we presented a versatile computation of the overall GWB arising from different astrophysical populations, with the goal of assessing how much it might affect the performance of space-based interferometers across the low frequency GW band ($\mu$Hz to 1Hz). To this end, we examined the four detectors LISA, $\mu$Ares, AMIGO, and DO, and we evaluated the GWB produced by MBHBs, EMRIs, SOBBHs, GBs and extragalactic DWDs.

In all four detectors, the signal from GBs acts as a foreground: it has little impact on DO but becomes increasingly significant in LISA and even more in $\mu$Ares and AMIGO. Most binaries with the highest orbital frequencies are resolved, producing an exponential cut-off in the signal, which occurs at a similar frequency across all detectors.

We observe that $\mu$Ares manages to resolve the majority of the detectable MBHBs already at the iteration $i=1$ of our code, since they are very loud signals that emerge starkly from the collective signal coming from the whole MBHB population. At each following step, additional sources are identified and extracted, repeatedly lowering the background level at frequencies higher than $10^{-5}$. Nonetheless, the residual signal from MBHBs and GBs overwhelms the instrumental PSD and shift up the detector sensitivity of $\sim 2-3$ orders of magnitude, dramatically degrading its performance at frequencies below $\sim \SI{1}{\milli \hertz}$. In spite of that, the predicted new effective sensitivity curve, which accounts for the overall astrophysical GWB, is enhanced in respect to LISA in the $\mu$-Hz window, thereby enabling the potential detection of new GW sources.

DO is particularly effective in reducing the residual background signal for EMRIs and SOBBHs. In both cases the observatory iteratively resolves more and more sources and succeeds in decreasing the GWB level to about one order of magnitude below the noise curve at frequencies of the order of $\SI{0.1}{\hertz}$. This could be a critical factor in the detection of subdominant stochastic GWB signals of cosmological origin. However, it is important to emphasize that our methodology assumes perfect source identification and subtraction, meaning that the waveform removed from the background uses the known true parameters. This is not realistic, since noise introduces errors in the parameter recovery, leading to imperfect source subtraction and spurious residuals in the data.
A more accurate estimate of the total GWB is given by the sum of the collective signal produced by undetectable events and the contribution from imperfect subtraction of the resolved sources \cite{Berti_2023}. This may prevent the actual reduction of the astrophysical background much below the detector noise limit, hence complicating the measurement of secondary GWBs of cosmological origin.

The design of AMIGO provides the same frequency band as LISA, but with increased sensitivity. However, this improvement is partially spoiled by the Galactic foreground, since AMIGO is not able to suppress it more effectively than LISA. In addition, above $\SI{3}{\milli \hertz}$, the GWB from extragalactic DWDs becomes dominant over the instrumental sensitivity; this contribution cannot be reduced since it is due to the sum of a very large amount of extremely faint sources that cannot be individually resolved, regardless of the detector performance. This leads to a limited gain in sensitivity only below $\sim \SI{10^{-4}}{\hertz}$ and above $\sim \SI{10^{-2}}{\hertz}$.

\subsection{Individual source detection}

With a realistic estimate of the residual GWB we can properly assess the individual source detection performance of each mission.

\paragraph*{$\mu$Ares.}
Despite being dominated by the MBHB GWB, the low-frequency sensitivity of $\mu$Ares allows for a more prolonged observation of the inspiral phase of MBHBs. Unlike the other missions, the resolved systems of $\mu$Ares are not limited to binaries approaching the coalescence phase within the observation time: it succeeds also in the detection of systems that are up to a few hundred years distant from the merger. These slowly orbiting MBHs are separated by distances on the order of $10^{-4}$pc and are found at very low redshift. Their observation can yield significant insights into the formation and evolution of binary systems of MBHs in the center of galaxies. The physics of MBH pairing is not yet fully understood, with multiple mechanisms that interplay, as stellar hardening, gas torque from a circumbinary disk or triple interactions \cite{Sesana_Khan_2015, Bortolas_2018, Roedig_2012, Bonetti_MBHB}. Detecting MBHBs during their early adiabatic inspiral - when the interaction with the environment can still leave distinctive signatures - could enhance our understanding of these processes \cite{Tang_2017, Spadaro_2025}. In addition, when embedded in gaseous environments, MBHBs at separation of $a/M\sim100$ are predicted to be fully coupled with gas, resulting in the production of a bright EM signal \cite{Mangiagli_2022,Franchini_2024}. This feature would make them ideal multimessenger targets.
This detector is able to identify the merging systems earlier in their inspiral, from a few years to several months before the coalescence. This results in a longer observation of the signal, which could allow for a precise sky localization before the merger. The latter is strongly related to the probability of finding an EM counterpart for the GW signal: resulting in a multimessenger candidate. The earlier we start to accurately identify the source position, the more we can continuously monitor it with instruments, such as the Vera Rubin Observatory or SKA, in order to search for a quasi-periodic EM signal that tracks the GW phase \cite{Zoltan_2017}. The detection of this signal is greatly promising for the most massive sources detected by $\mu$Ares, as they are found at low redshift, leading to a brighter EM emission.

\paragraph*{DO.} This detector would sit in the gap between LISA and ground based interferometers, covering a frequency band largely clean from astrophysical backgrounds. In fact, besides the irreducible extragalactic DWD GWB, other astrophysical sources can be deterministically matched, keeping the unresolved GWB below the detector sensitivity at $f>0.1$ Hz. Considering the MBHB population, in the LS scenario, the less massive systems merge in the deci-Hz band, and DO will resolve all of them. This will allow us to infer the mass distribution of MBHs as a function of the redshift, separating different seed formation channels \cite{2011PhRvD..83d4036S} and MBH evolution, such as super-Eddington flows \cite{Madau_2014} and binary mergers.
The above considerations can be applied also to the SOBBH population. By measuring the source parameters of tens of thousands of sources, we could reconstruct the properties of the population of these systems across redshifts in relation to the star formation history and metallicity evolution in the Universe \cite{2014ARA&A..52..415M}. We could also discriminate between separate formation mechanisms involving isolated binaries and dynamically assembled systems \cite{2016PhRvD..94f4020N,Breivik_2016}. Finally, regarding the EMRI population, DO manages to resolve a considerable number of systems, characterized by a less massive central MBH and a higher value of eccentricity compared to LISA. This will probe the low mass end of the quiescent MBH mass function with unprecedented precision, will provide new insights on the dynamics of dense nuclei and will shed light on the role played by accretion of compact objects in the early growth of MBHs. 

\paragraph*{AMIGO.} Designed as a ten-fold improvement of the LISA sensitivity, AMIGO primarily targets the milli-Hz band, with a scientific return that takes elements from $\mu$Ares and 
DO. Considering MBHBs, it manages to resolve most low-mass merging MBHBs also in the LS scenario, allowing a better recosntruction of their cosmic history compared to LISA. It also manages to detect few MBHBs still far from coalescence, albeit much less than $\mu$Ares, thus also resulting in appealing to the multimessenger front. About a thousand SOBBHs will be detected up to $z\approx 1$. The high number of detections will allow to trace back different formation scenarios from potential features in the measured parameter distribution (masses, eccentricity, spins), although the restriction to low redshift will make it hard to connect the compact object merger rate to the cosmic star formation history. Turning to EMRIs, the capabilities of AMIGO are comparable to DO, allowing the detection of thousands of systems to $z\approx 4$. This will help in reconstructing the low mass end of the MBH mass function, assessing the role of EMRIs in MBH growth and understanding the dynamics of dense nuclei. Finally, we note that, despite the ten-fold improvement in sensitivity, AMIGO will not be able to pierce through the galactic and extragalactic DWD foregrounds, which will limit its sensitivity in the $10^{-4}-10^{-2}$Hz range. In this sense, the design of LISA is nearly optimal for the detection of deterministic signals around the milli-Hz region.

\subsection{Caveats}
\label{sec:caveats}

Our calculations rely on a number of assumptions and simplifications in the treatment of the astrophysical population, in the modeling of the GW signals and in the implementation of the detector response. We briefly enumerate them here, justify our choices, and discuss their impact on the results.

\subsubsection{Astrophysical populations and waveform modeling}

On the astrophysical population side, we relied on fiducial models for MBHBs, EMRIs, SOBBHs, Galactic binaries (GBs) and extragalactic DWDs published in the literature. In absence of a solid, fiducial model, we neglected any IMRI population, which might further contribute to the overall GWB. We stress, however that the modularity of our approach makes it trivial to add further populations.

We assumed all binaries to be composed by non-spinning objects in quasi-circular orbits, except for EMRIs which can be very eccentric. Although the circular approximation is justified by GW-driven circularization, this might not hold far from merger, where binaries are still likely to carry memory of their formation channel. This is particularly true for SOBBHs at milli-Hz frequencies and for MBHBs at micro-Hz frequencies. Highly eccentric binaries evolve much faster, with the net result that fewer sources are active at any time, diminishing the GWB at low frequencies \citep[i.e. prior to circularization,][]{2013CQGra..30v4014S,2018MNRAS.481.4775D}. 
Ignoring spins in our calculation is a safer assumption, since they mostly affect the late phases of the inspiral and, contrary to eccentricity, do not drastically change the strain produced by each source. The SNR of spinning binaries merging in band might be significantly higher compared to non-spinning systems, affecting the number of sources that can be individually identified and subtracted from the data stream. This might be important for MBHBs. It should be noticed, however, that in the HS model, all merging binaries are individually resolved already under the non-spinning approximation, whereas for the LS model, faint merging binaries contributing to the GWB for LISA and $\mu$Ares merge way out of band anyway, and including spins in their SNR calculation would not affect their detectability.

All GW strains, except for MBHBs are modeled at the quadrupolar order following Eqs.~\eqref{eq:hc_code_ecc} and \eqref{eq:hc_code}. Although more sophisticated waveforms might produce slightly different SNR values, this is expected to produce only minor changes in the overall GWB computation. This was demonstrated for EMRIs in \cite{Pozzoli_EMRI}, who found that the level of EMRI GWB calculated by employing the sophisticated FEW waveform package \cite{2021PhRvD.104f4047K} was consistent with the one computed in \cite{Bonetti_EMRI}, which employed the simplified Kludge waveforms of \cite{2004PhRvD..69h2005B}.

Finally, although we use the full PhenomC waveform to compute the SNR of individual MBHBs, we just consider the inspiral according to Eq.~\eqref{eq:hc_code} when computing the contribution to each system to the stochastic GWB. This is justified by the fact that in the HS model all merging binaries are resolvable by all detectors, and the ones contributing to the GWB are seen only in their inspiral. This is not true for the LS model, in which merging binaries with $M\lesssim10^3$M$_{\odot}$ are generally not resolved, thus contributing to the overall GWB. It should be noted, however, that the merger of these light binaries occur at frequencies $\gtrsim 1$ Hz, beyond the band of interest for this study. Moreover, the residual GWB from MBHBs at those frequencies is buried in the one produced by the far more abundant SOBBHs.
 
\subsubsection{Detector response and observation treatment}

We simplified the treatment of the detection process by assuming no detector motion, sky averaged sensitivity and inclination-polarization averaged GW strain of each individual source. Although these assumptions are admittedly simplistic, it was again demonstrated in \cite{Pozzoli_EMRI} that they are sufficient for a trustworthy estimate of the GWB. This is because, while the properties (and SNR) of each individual source do depend on its inclination-polarization and on the detector motion, these effects are smoothed out when averaging over the many systems contributing to the GWB.

Finally, to identify the level of irreducible GWB, we assumed perfect subtraction of all individually resolvable sources, based on a purely SNR criterion. The latter ignores effects as signal overlap in the same frequency bin, which is relevant for Galactic binaries. This can prevent our ability to gain enough information to successfully extract the signal, since at least a few frequency bins are needed to fully characterize the 8 source parameters. However, as a rough estimate, there are $\sim 10^4$ resolved sources at frequencies $\sim 10^{-3}$Hz distributed across $\sim 10^5$ frequency bins (width $\sim 1 / T_{obs}$), suggesting that source density should have minimal impact in the resolvability. This aligns with a previous study that accounted for this argument \cite{2006PhRvD..73l2001T}, showing that including additional criterion on the source density has a smaller influence on the detectability than the noise contribution from the unresolved background, since the latter dominates the high density regions. Noticeable differences can emerge when considering a lower SNR threshold (e.g. 5) or Galactic models predicting a very high number of sources. Not accounting for the imperfect removal of the resolved sources is also an optimistic assumption, but we remark that proper source subtraction is still an open problem in GW data analysis and its impact on the residual GWB should be evaluated in a global fit context, which is beyond the scope of this work.

\section{Conclusions}
This work provides a preliminary overview of which types of astrophysical systems are most promising for future space-based observatories, carefully accounting for the stochastic noise of astrophysical origin. Despite the significant impact of the latter on $\mu$Ares’s sensitivity, this mission proposal would open a unique window for multimessenger astrophysics. Unlike the other missions, which are limited to the detection of merging MBHBs, $\mu$Ares can observe these binaries during their inspiral phase, in a regime with an enhanced chance of EM emission. In addition, it detects the same merging systems observed by the other missions, but at an earlier stage: this would enable longer observations of the signal and an accurate sky localization well ahead the merger, essential for joint GW and EM studies. Conversely, the main strength of a GW interferometer operating in the deci-Hz band lies in its ability to probe sources out to high redshift. One of the targets of DO  are MBHBs at $z \sim 15-20$, predicted by the LS model for MBH formation and evolution. DO is also able to extend the redshift horizon of SOBBHs and EMRIs up to $z \approx 10$ and $z \approx 4.5$, respectively, based on our synthetic catalogs. Finally, since this mission design is the least affected by astrophysical backgrounds, it could provide a clearer view of stochastic GWBs of cosmological origin. Some of these scientific cases are also relevant for AMIGO, although to a lesser extent, since its sensitivity is limited by the foreground produced by both Galactic and extragalactic DWDs. 

As detailed in Sec.~\ref{sec:caveats}, our methodology relies on a number of assumptions and simplifications. Future extensions of this study should include a more accurate calculation for the subtraction of the loudest sources: the contribution from possible imperfect removal has to be taken into consideration in order to have a correct evaluation of the capability of the detectors in cleaning the foreground signal. In addition, our analysis does not account for the possible, at least partial, subtraction of the astrophysical foregrounds due to their departure from isotropy, Gaussianity and stationarity.
For example, the Galactic DWD foreground can be identified due to its anisotropic modulation, which traces the spatial distribution of the Milky Way \cite{2025PhRvD.111f3005P}. It is also plausible that a significant contribution to the GWB foreground of EMRIs and MBHBs comes from a relatively small number of slightly subthreshold event, implying significant non-Gaussianity and non-stationarity of the resulting signal. This has been demonstrated to be the case for EMRIs in \cite{2025PhRvD.111j3047P}. Bayesian methods for spectral separability of anisotropic and isotropic GWBs are currently being investigated in the context of LISA \cite{Criswell_2025}, in order to distinguish the contributions of different astrophysical populations and potentially measure a cosmological background.

\section*{Acknowledgments}
We thank members of the Gravitational-Wave Space 2050 Working Group for their useful discussions.
This project has received financial support from the CNRS through the MITI interdisciplinary programs through its exploratory research program. A.P. acknowledge support by the French Agence Nationale de la Recherche. M.B. acknowledges support from the Italian Ministry for Universities and Research (MUR) program “Dipartimenti di Eccellenza 2023-2027”, within the framework of the activities of the Centro Bicocca di Cosmologia Quantitativa (BiCoQ). A.S. acknowledges financial support provided under the European Union’s H2020 ERC Consolidator Grant ``Binary Massive Black Hole Astrophysics" (B Massive, Grant Agreement: 818691) and Advanced Grant ``PINGU'' (Grant Agreement: 101142079).